\documentclass{ieeeaccess}
\usepackage{cite}
\usepackage{amsmath,amssymb,amsfonts,braket}
\usepackage{algorithmic}
\usepackage{graphicx}
\usepackage{textcomp}
\usepackage{subfig}   
\def\BibTeX{{\rm B\kern-.05em{\sc i\kern-.025em b}\kern-.08em
    T\kern-.1667em\lower.7ex\hbox{E}\kern-.125emX}}
\usepackage[utf8x]{inputenc}

\begin{document}
\history{Date of publication xxxx 00, 0000, date of current version xxxx 00, 0000.}
\doi{10.xxxx/TQE.2024.DOI}

\title{Long-Range Entangled Quantum Noise Radar Over Order of Kilometer}
\author{\uppercase{H. Allahverdi}\authorrefmark{1,}\authorrefmark{2} and
\uppercase{Ali Motazedifard\authorrefmark{1,}\authorrefmark{3,}\authorrefmark{4}}}
\address[1]{Quantum Remote Sensing Lab, Quantum Metrology Group, Iranian Center for Quantum Technologies (ICQT), Tehran, Tehran 15998-14713, Iran}
\address[2]{Laser and Plasma Research Institute, Shahid-Beheshti University, Tehran, Tehran 19839-69411, Iran}
\address[3]{Free-Space Quantum communication Group, Iranian Center for Quantum Technologies (ICQT), Tehran, Tehran 15998-14713, Iran}
\address[4]{Quantum Optics Group, Department of Physics, University of Isfahan, Hezar-Jerib, Isfahan 81746-73441, Iran}

\markboth
{Allahverdi \headeretal: Long-Range Entangled Quantum Noise Radar Over Order of Kilometer}
{Allahverdi \headeretal: Long-Range Entangled Quantum Noise Radar Over Order of Kilometer}

\corresp{Corresponding author: Ali Motazedifard (email: motazedifard.ali@gmail.com).}

\begin{abstract}
In this paper, 
an explicit expression for the maximum detection range of an entangled quantum two-mode squeezed (QTMS) radar, in which a two-mode squeezed vacuum state of microwave electromagnetic fields is used, have been derived by considering both the quantum properties of the entangled microwave fields and radar parameters. By comparing this equation with that of traditional radars, we showed that one can though a QTMS radar as a traditional radar with a reduced threshold signal-to-noise ratio.  
By discussing the current limitations, it has been shown that the critical parameter to have both simultaneous quantum advantage and substantial radar range is increasing the bandwidth of the generated output signal in the quantum entangled source. It has been shown that by considering the current feasible system parameters, it is possible to implement a QTMS radar with maximum detection range up to the order of $2\mathrm{km}$, which is suitable for recognizing small unmanned aerial vehicles in urban distances. 
Moreover, based on the false alarm rate, we introduce two classes of early alarm and track QTMS radars.
The present approach can be generalized to other quantum radars with different types of quantum sources like electro-opto-mechanical sources, and also may shed new light on investigating the quantum radar system toward practical applications.
Finally, we have discussed the potential outlooks to improve and develop the quantum entangled radar systems to be practical from the engineering point of view.
\end{abstract}


\titlepgskip=-15pt

\maketitle

\section{Introduction}\label{sec.introduction}


\Figure[t!](topskip=0pt, botskip=0pt, midskip=0pt){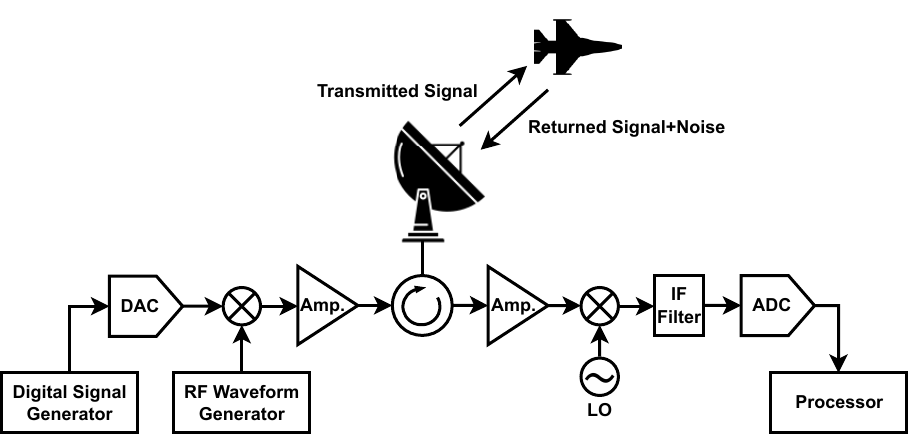}
{Schematic diagram of a mono-static direct detection radar system. Amp.: Amplifier; DAC: Digital to analog converter; ADC: Analog to digital converter; LO: Local oscillator.\label{fig1}}

\PARstart{D}{etection} and characterization of nearby objects aided by microwave electromagnetic (EM) fields, which usually refers to RADAR (Radio frequency (RF) Detection And Ranging), originated to the pioneering works of German physicist, H. Hertz \cite{historyRadar}. 
The operation principle of a radar is so simple: it transmits a microwave EM signal to the target's region, and then analyzes the returned signal to determine whether the target is present or not, and estimates its characteristics such as image, range, azimuth, and velocity. Nowadays, Radars have found applications in different areas from biology \cite{radarBook2,wirelessBioRadar} to astronautics \cite{cubeSat}, navigation \cite{radarnavigationBook}, aviation safety \cite{radarNetworks}, and also smart wearable devices \cite{smart}. 

Traditional radars, which we call direct detection radars, analyze the power of the received signal to identify the target, i.e., the received signal power $P_r$ is the detector function of the detection system \cite{skolnikBook}. However, their operation is easily failed by unwanted signals like environmental thermal noise and other sources of microwave signals at the same frequency. To facilitate the recognition of unwanted signals from radar signals that bounced off from the target, engineers proposed to keep a record of the transmitted signal as a reference, and then compare the received signal with this reference (idler) \cite{narayananNoiseRadarDesign,lukinNoiseRadarTechnology,noiseRadarOverview,efficientProcessingNoiseRadar2018}. In these radars, usually known as noise radars, the correlation between the received signal and the idler is the detector function of the detection system. As the correlation between the transmitted signal and the retained idler is greater, identification of the target becomes more and more accurate.
The question that immediately arises is ``\textit{what is the ultimate limit of the correlation between the signal and idler in noise radars}?" From the classical point of view, the highest correlation is achieved when the transmitted signal and the retained idler have exactly the same waveforms \cite{narayananNoiseRadarDesign}. 
Here is where the quantum mechanics plays an important role: if a pair of entangled EM fields have been employed as the signal and the idler, then more correlations in comparison to the classically correlated signals can be extracted through, and thus, the precision of the target detection increases substantially \cite{torrome2023advances,Shapiro2020story,Pirandola2018advance}. This is the main idea of quantum illumination (QI), which has been proposed by S. Lloyd in 2008 \cite{quantumilluminationSethLloyd2008}. The first experimental realization of QI was performed in the optical domain using bulk nonlinear optics by E.D. Lopaeva \textit{et.al.} in 2013 \cite{lopaevaQuantumIllumination2013}. They showed that the QI system realizes a 6dB advantage in the error-probability exponent over the optimum classical system. The first proposal for QI in microwave domain was proposed by Barzanjeh \textit{et.al.} in 2015 \cite{barzanjehQIlluminationOMS2015}, in which they suggested to utilize an electro-optomechanical system as a source of entangled microwave-optical photons. 
Beyond the advantages of the QI, it requires a \textit{joint measurement} of the received signal and the retained idler which is impossible when the arrival time of the received signal at the receiver is vague. 
Because of this, the microwave QI system proposed in Ref.~\cite{barzanjehQIlluminationOMS2015} seems to be impractical for radar applications due to the unknown target's range. However, after that in 2018, the first microwave QI-inspired radar which did not need joint measurement, has been demonstrated experimentally by B. Balaji \textit{et.al.} \cite{wilsonMicrowaveQuantumRadar2018,snakeBalaji2018}. In this system, the so-called QTMS radar, a Josephson Parametric Amplifier (JPA) which is cooled down below 10mK by a cryostat was used as a source of entangled microwave signals.

In recent years, many theoretical and experimental papers have been published on the optical QI \cite{helmy40dBQuLiDARNature2022,helmy100dBQuLiDARNature2023,quantumDopplerLidar2022,jeffers1QuantumRangingJamming,jeffers2QuantumRangingJamming,quantumLidarFMCW,quantumSecuredLiDAR,twoPhotonLiDAR2022,jeffers3MimicQillumination,karsa2020quantum} and also
on the microwave QTMS radars \cite{barzanjehExperimentQuantumRadar2020,balajiCorrelationExperiment,vitaliQradarExperiment2023,vitaliQradarExperiment2022TechRxiv,vitaliQradarExperiment2021Conference} to illustrate the advantages of them with respect to the classical analogous with the same features. 
The majority of the works on microwave QI are focused on the effect of entanglement on signal-idler correlation enhancement, and the effect of this enhancement on performance of the system in terms of the receiver operation characteristic (ROC) curve and the detection signal-to-noise-ratio (SNR) which are not sufficient for analyzing the performance of these systems in long-range applications in reality. 
A standard approach to determine how far a QTMS radar is capable of detecting a target is to obtain a range equation that determines the signal-idler correlation in terms of the target range. B. Balaji \textit{et. al}. in Ref.~\cite{balaji2022PerformancePrediction} derived an explicit expression for Pearson correlation coefficient $\rho$ in terms of the target range $R$ for noise radars. However, they did not considered the atmospheric absorption, and consequently, their result can not be applied for real-world situations notably in long-range where the target is far away and atmospheric absorption of EM signals is significant. 
Furthermore, they have not considered the detection time period in their analyzes. 
In fact, beyond the fundamental importance of developing the QTMS radar idea, there is a vital question that arises from the investigators and engineers: can QTMS radars really work in long-range applications out of the Lab or not?! 

Up to where we know there is no clear calculative and precise answer to tackle this question. 
Some authors evaluated the maximum detection range of quantum radars by utilizing the classical approach that obtained for direct-detection radars \cite{rangeEquation2023,bischeltsrieder2024engineering}, in which the maximum range is derived from the classical radar detection condition in which the received signal power is considered as detection function. The applied approach is not theoretically complete because in quantum radars the correlation between the signal and the idler is the detection function, not the signal power. Moreover, none of these works take into account atmospheric absorption which is significant at long ranges and practical application.

In this paper, based on the above-mentioned issues, we are motivated to present an analytical approach
to answer this ambiguity in quantum radar context to tackle questions such as: \textit{what is the maximum detection range of a QTMS radar}?\textit{ Is a long-range QTMS radar really possible}? 
To answer these questions clearly, 
we analytically derive an explicit expression for the maximum detection range $R_{\mathrm{max}}$ for CW noise radars by considering the atmospheric absorption, detection time, and bandwidth using a similar approach in Ref.~\cite{balaji2022PerformancePrediction}. Moreover, we introduce the idea of single microwave photon (SMP) Radar by discussing on applying the quantum antennas or single microwave photon detectors (SMPDs).
By comparing to the classical direct detection radars, we show that a noise radar can be seen analogously as an enhanced direct detection radar. 
We then compare the performance of QTMS radar with an ideal classical noise radar in terms of the maximum detection range, and surprisingly, show that using the current feasible experiments, the long-range QTMS radar in order of km is achievable in real scenarios. Here, we present an analytical expression for quantum range equation, and interestingly discuss how should overcome the present limitations and challenges in order to increase the maximum range, for example by enhancing the entanglement, or using SMPDs, and more importantly the increase of the signal-bandwidth in the microwave quantum entangled source. The latter leads to a strong quantum advantage via the decrease of the photon per mode which implies the long-range feasibility as recently the authors in Refs.~\cite{vitaliQradarExperiment2023,vitaliQradarExperiment2022TechRxiv} have argued. Furthermore, we compare the SNR of a quantum entangled radar with both the classical direct detection and noise radars.

The paper is organized as follows. In Sec.~\ref{sec2.DirectDetectionRadars}, we review the range equation in conventional classical direct detection radar, and derive its correspond analytical expression for maximum detectable range. We then introduce the noise radars in Sec.~\ref{sec3.NoiseRdar}. Utilizing the results obtained in Sec.~\ref{sec3.NoiseRdar}, we compare the performance of the QTMS radars and classical noise radars in Sec.~\ref{sec4.QuantumNoiseRadar}. Experimental discussion on the feasibility and comparison as well as addressing challenges is provided in Sec.~\ref{sec5.experiment}. Finally, conclusion remarks and outlooks for future works are discussed in Sec.~\ref{sec6.Results and Discussion}.

\section{Direct Detection Radars} \label{sec2.DirectDetectionRadars}

In this section, to compare with the entangled-assisted quantum radar we first review the operation principle of direct detection radar systems and derive an analytical expression for the maximum range over which the CW-radar system can detect the target for both monostatic and bistatic cases. Finally, we introduce the single-photon Radar.

The schematic of a direct detection radar is illustrated in Fig.~(\ref{fig1}), where a digital signal generator produces an intermediate frequency (IF) signal that is converted to an analog one via a digital-to-analog converter (DAC) and up-converted to a RF frequency by mixing with a RF waveform. The generated RF signal then is amplified and sent to the radar antenna to be transmitted toward the target region. 
The transmitted microwave signal (with power $P_t$) is attenuated in the atmosphere and after interaction with the target is partially reflected toward the radar receiver. 
At the receiver, the received signal (with power $P_r$) and the environmental noises are amplified and then down-converted to an IF or based-band (BB) frequency by mixing with a RF local oscillator (LO). After that, it passes through an IF bandpass filter (with a bandwidth of $B_{\mathrm{IF}}$) and is digitized using a fast analog-to-digital converter (ADC), and thus finally sent to a processor such as FPGA for processing and interpretation. 
The returned signal power $P_r$ is related to the transmitted signal power $P_t$ through the standard classical range equation 
\cite{skolnikBook}
\begin{align}
& P_r=\eta (R) P_t , \label{power1}
\end{align}
in which $ \eta(R) $ is the transfer function of the transmitter-target-receiver channel and given by \cite{skolnikBook}
\begin{align}
& \eta(R)= \dfrac{\sigma G A_e}{(4\pi)^2 R^4} F(R)^2, \label{eta1}
\end{align}
where $ R $ is the range to the target, $ \sigma  $ is the radar cross-section (RCS), $ G $ is the antenna gain, $ A_e $ is the effective aperture of the receiver antenna and related to the physical antenna aperture $ A $ via $ A_e=\epsilon_a A $ with $ \epsilon_a $ being the antenna aperture efficiency, and also $ F(R)=10^{-\gamma R/10}  $ is the form-factor describing the attenuation of the radar signal due to atmospheric absorption or loss with $ \gamma $ being the atmosphere absorption coefficient in the units of $\mathrm{dB/m} $.

In addition to the returned signal, there is an unwanted contribution due to the environmental noise with power $ P_n $ in the total power received by the receiver. If the process of target detection takes place over $M$ measurements, then the effective signal-to-noise ratio, $ {\mathrm{{SNR}_{eff}}} $, is given by \cite{skolnikBook}
\begin{align}
& \mathrm{{SNR}_{eff}^{dr}}=\dfrac{MP_r}{P_n }= M \eta(R)  \dfrac{P_t}{P_n}. \label{SNR1}
\end{align}
It is worth noting that the number of measurements $ M $ is determined by the integration time $ \tau_{\mathrm{int}} $  and detection bandwidth $B_{\mathrm{det}}$ which itself determines by the IF filter bandwidth $B_{\mathrm{IF}}$ at the receiver (i.e., $B_{\mathrm{det}}=B_{\mathrm{IF}}$), such that $ M=\tau_{\mathrm{int}}B_{\mathrm{det}} $.

\subsection{Detection Condition in Direct Detection Radars}

\begin{figure*}
	\centering
	\subfloat[]
	{\includegraphics[width=8cm]{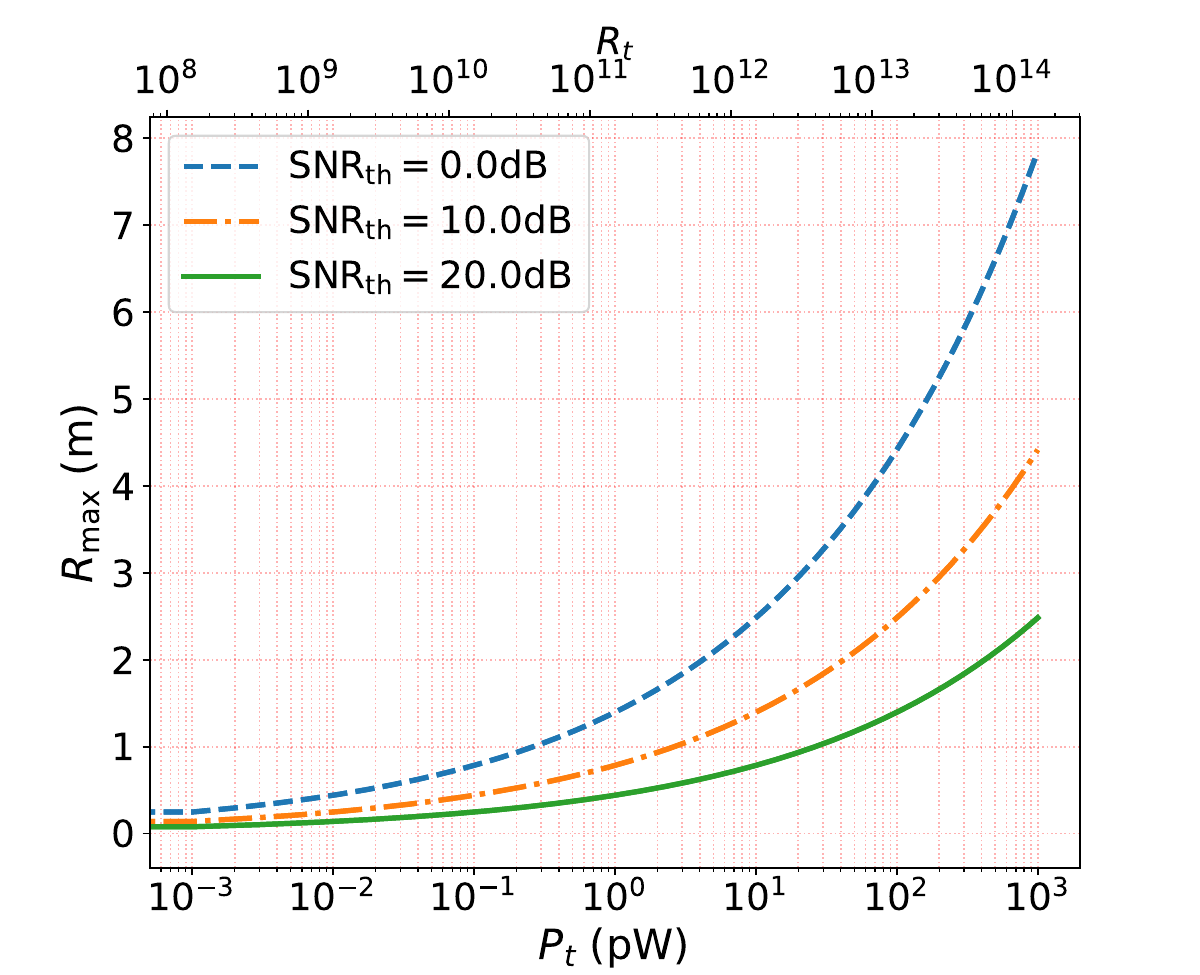}}\label{fig2a}
	\subfloat[]
	{\includegraphics[width=8cm]{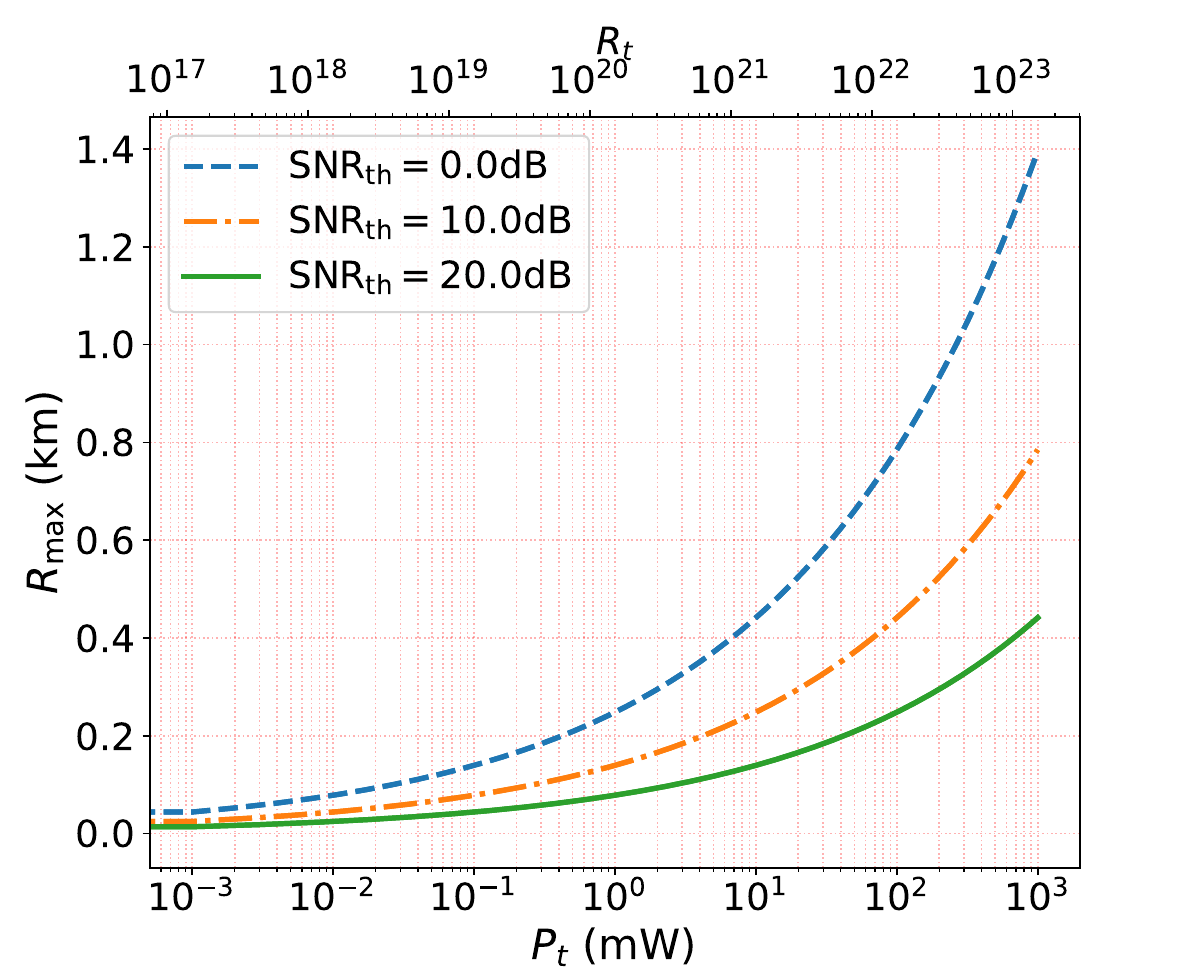}} \label{fig2b}
	\centering
	\caption{(Color online) Maximum detection range of a direct detection radar $R_{\mathrm{max}}^{\mathrm{dr}} $ with respect to the power of the transmitted signal $ P_t $ (lower horizontal axis) and the corresponding transmitted photon rate $R_t$ (upper horizontal axis) for different values of the threshold SNR (i.e., $ \mathrm{{SNR}_{th}} $), in (a) low-brightness, and (b) high-brightness regimes. Parameters considered in this plot are given in the table.~(\ref{table1}).  }
	\label{fig2}
\end{figure*}


A direct detection radar is able to resolve the target (if present) if the effective $\mathrm{SNR}$ to be greater than a conventional threshold ${\mathrm{{SNR}_{th}}}$ \cite{skolnikBook}, i.e., 
\begin{align}
& {\mathrm{{SNR}_{eff}^{dr}}} \ge {\mathrm{{SNR}_{th}}}. \label{SNR_cond}
\end{align}
It is worthwhile to mention that ${\mathrm{{SNR}_{th}}}$ refers to the minimum required power of the returned signal, $P_{r,\mathrm{min}}$, or minimum detectable signal (MDS), for which the radar system enables to recognize the target, so that ${\mathrm{{SNR}_{th}}}=P_{r,\mathrm{min}}/P_n$. 
For instance, for a typical radar system with $\mathrm{{SNR}_{th}}=10~dB$, it means that the power of the returned signal $P_r$ should be at least $10~\mathrm{dB}$ higher than the environmental noise power $P_n$ to radar system be able to identify the target, i.e., $P_r~(\mathrm{dB})\ge P_n~(\mathrm{dB}) + 10~\mathrm{dB}$, or equivalently, $P_r~({\mathrm{SI}})\ge 10\times P_n~({\mathrm{SI}})$.

By applying \eqref{SNR1} into the above equation, we found that
\begin{align} \label{rangeEq_1}
&  R 10^{(\gamma R/20)} \le (\dfrac{M}{\mathrm{{SNR}_{th}}})^{1/4} R_c^{\mathrm{dr}}   , 
\end{align}
in which $ R_c^{\mathrm{dr}} $ is the characteristic-range of the direct detection radar system and defined as
\begin{align} \label{r_c_direct} 
& R_c^{\mathrm{dr}} \equiv \left( \dfrac{\sigma G A_e P_t}{(4\pi)^2 P_n} \right)^{1/4}.  
\end{align}
$R_c^{\mathrm{dr}}$ is a range for which the total power at the receiver for in-vacuo signal propagation ($\gamma=0 $) is 3dB  above the noise level, or equivalently, twice the noise power $ P_n $. 
Alternatively, it can be interpreted as a range for which the SNR  at the receiver equals 1 (or 0 dB ). 
The characteristic range depends on the system parameters, transmitted signal power $P_t$ and the environmental thermal noise power $P_n$, which given by \cite{skolnikBook}
\begin{align} \label{n_thermal} 
& P_n=N_b h f B_{\mathrm{det}},
\end{align}
in which $N_b=[\exp(hf/k_BT)-1]^{-1}$ is the mean photon number of the background thermal noise, $f$ is the signal frequency, $h=6.63\times10^{-34}~\mathrm{j.s}$ is the Planck constant, $k_B=1.38\times10^{-23}~\mathrm{j.K}$ is the Boltzmann constant, and $T$ refers to the temperature. For instance, for $f=10 ~\mathrm{GHz}$, $T=300~ \mathrm{K}$, and $B_{\mathrm{det}}=200~\mathrm{kHz}$, we have $N_b\simeq624$ and $P_n\simeq-120.82~\mathrm{dBm}$.

\subsection{Maximum Detection Range in Direct Detection Radars}
For a given transmitted power $ P_t $, the equality in Eq.~(\ref{rangeEq_1}) holds for the maximum detection range of the direct detection radar, $ R_{\mathrm{max}}^{\mathrm{dr}} $. Therefore, using \eqref{rangeEq_1} we get to
\begin{align} 
&  R_{\mathrm{max}}^{\mathrm{dr}}  \mathrm{exp}\big(0.115 \gamma R_{\mathrm{max}}^{\mathrm{dr}}\big) = (\dfrac{M}{\mathrm{{SNR}_{th}}})^{1/4}  R_c^{\mathrm{dr}},   \label{dr_range_eq_raw}
\end{align}
which implies that the maximum detection range $ R_{\mathrm{max}}^{\mathrm{dr}} $  increases by increasing $ R_c^{\mathrm{dr}} $. 
It is obvious that to have a radar with a longer range, a greater $ R_c^{\mathrm{dr}} $ is needed, which means a convenient set of system parameters must be designed. 
Equation \eqref{dr_range_eq_raw} can be analytically solved to obtain an explicit expression for $R_{\mathrm{max}}^{\mathrm{dr}} $ as follows
\begin{align} \label{rangeEqDirectFinal}
&  R_{\mathrm{max}}^{\mathrm{dr}} =  \dfrac{1}{0.115 \gamma} {\mathrm{ln}}\left[  1+ 0.115 \gamma \left(\dfrac{M}{{\mathrm{{SNR}_{th}}}}\right)^{1/4} R_c^{\mathrm{dr}} \right].
\end{align}

Figure \ref{fig2} shows the behavior of $R_{\mathrm{max}}^{\mathrm{dr}} $ with respect to the transmitted power $P_t$ (lower horizontal axis) and transmitted photon rate $R_t\equiv P_t/hf$ (upper horizontal axis) for feasible experimental parameters given in table.~(\ref{table1}). 
Here, it is worth reminding that $ R_t=N_t B $ where $ N_t $ and $ B $ are the mean transmitted signal photons and signal bandwidth, respectively.
To well-illustrate the magnitude of $R_{\mathrm{max}}^{\mathrm{dr}}$ for both extremely low powers and moderate to high powers, we divide the transmitted power into two regimes in Fig.~(\ref{fig2}): low- and high-brightness regimes. 
The high-brightness regime (see Fig.~\ref{fig2}(b)), in which $P_t>1~\mu\mathrm{W}$, is the usual regime in which most of the practical classical radars work, and the radar can identify targets in km-ranges in this regime. In the low-brightness regime (see Fig.~\ref{fig2}(a)), the transmitted power is lower than $1~\mathrm{nW}$, and the considered direct detection radar is only capable of identifying targets closer than 10 meters.
However, the importance of the low-brightness regime is that most advantages of quantum entangled radars lie in it. Therefore, we have plotted the maximum range of the direct detection radar in this regime for comparisons with CW noise radar and quantum entangled radar that will be introduced in Sec.~\ref{sec3.NoiseRdar} and Sec.~\ref{sec4.QuantumNoiseRadar}, respectively. 
According to Fig.~(\ref{fig2}), it is clear that $R_{\mathrm{max}}^{\mathrm{dr}} $ increases by increasing the transmitted power $ P_t $ in both regimes. This is because the received signal at the receiver $P_r$ is stronger for higher transmitted powers $P_t$ (see \eqref{power1}), and consequently, the detection range at which the received power equals the least required power for target detection, i.e., $P_{r,\mathrm{min}}$, increases. 
Finally, it can be concluded that in the low-brightness regime, the conventional radar system can operate no longer than a few meters, while it will be shown that a long-range quantum entangled noise radar is achievable in this regime.
Moreover, Fig.~(\ref{fig2}) illustrates the effect of $ {\mathrm{{SNR}_{th}}} $ on the maximum detection range. It is evident that for a given $ P_t $ and $ R_{\mathrm{max}}^{\mathrm{dr}} $, the maximum detection range increases substantially by lowering the $ {\mathrm{{SNR}_{th}}} $. 
As stated previously, ${\mathrm{{SNR}_{th}}}$ is obtained from the minimum returned power $P_{r,\mathrm{min}}$ required for the radar receiver to identify the target, which itself depends on the efficiency of the microwave receiver, i.e., its sensitivity.

\begin{table}
	\caption{Typical Parameters for a direct detection radar system.}
	\label{table1}
	\setlength{\tabcolsep}{3pt}
	\centering
	\begin{tabular}{|p{10pt}|p{45pt}|p{45pt}|p{45pt}|}
		\hline
		No.& 
		Parameter& 
		Unit & Value \\
		\hline
		1 & $ f $ & $\mathrm{GHz}$ & 10  \\
		2 & $ \sigma $ & $\mathrm{m}^2$ & 0.5  \\
		3 & G & dB &  15 \\
		4 & $ \epsilon_a $ & - & 0.5 \\
		5 & A & $\mathrm{m}^2 $ & 0.01$ \pi $ \\
		6 & $T$ & $\mathrm{K} $ & 300 \\
		7 & $ \tau_{\mathrm{int}} $ & ms & 100 \\
		8 & $ B_{\mathrm{det}} $ & kHz & 200 \\
		9 & $B$	& $\mathrm{MHz}$ & 100 \\
		10 & $ \gamma $ & dB/km & 0.007 \cite{ITUR}   \\
		\hline
	\end{tabular}
	\label{tabel1}
\end{table}

\subsection{Single-Photon Radar}

As is clear, the threshold-SNR of the radar receiver has non-negligible effect on the maximum range of a direct detection radar in the low-brightness regime.
Development of microwave detectors with a substantially improved detection sensitivity yields an enhanced $ {\mathrm{{SNR}_{th}}} $, which consequently increases the range of a conventional direct detection Radars, $R_{\mathrm{max}}^{\mathrm{dr}} $, notably in the low-brightness regime where current systems cannot have an applicable range. 
However, the implementation of single photon detection in the GHz band is challenging due to the 5-order smaller photon energy, but fortunately, recent progress in superconducting qubits or bolometers based on the irreversible transfer of photons to the excited state of a transmon qubit via a 4-wave mixing process yields the-state-of-the-art SMPD \cite{singleMicrowaveDetector1} with the astonishing sensitivity of $ 10^{-22}~ \mathrm{W}/\sqrt{\mathrm{Hz}} $ around frequency of 7GHz with tunability of $400~\mathrm{MHz}$ with $1~\mathrm{MHz}$-bandwidth, $ 45\% $-efficiency, and dark count of $85~\mathrm{cps}$ at $190~\mathrm{mK}$ temperature which enables to detect a weak RF-signal with power in order of $ -190~\mathrm{dBm} $ which is equivalent to ${\mathrm{{SNR}_{th}}}=-54~\mathrm{dB}$. 
Therefore, the emergence of SMPDs immediately addresses a new type of quantum direct detection radar that we name it ``Single-Photon Radar" analogous to the optical counterpart, i.e., single-photon LiDARs \cite{degnan2016,singlePhotonLiDARsPawlikowska,liDAR200km2021,singlePhotonLiDARSubMerge2023}, for operating in microwave low-brightness regime. Therefore, by considering the MDS achievable in the above-mentioned receivers, one can assume that in a single-photon radar, a threshold-$ \mathrm{SNR}$ in order of $-54~\mathrm{dB}$ is feasible. Thus, interestingly it yields to have a direct detection single-photon radar with maximum detection range in order of few hundred metes in the low-brightness regime. It means that about 2-order of magnitude enhancement in the radar range is achievable in the low-brightness regime if one can replace a SMPD, as a quantum antenna, by the current classical receiver antenna.

\section{Noise Radars} \label{sec3.NoiseRdar}

\begin{figure*}
	\includegraphics[width=14cm]{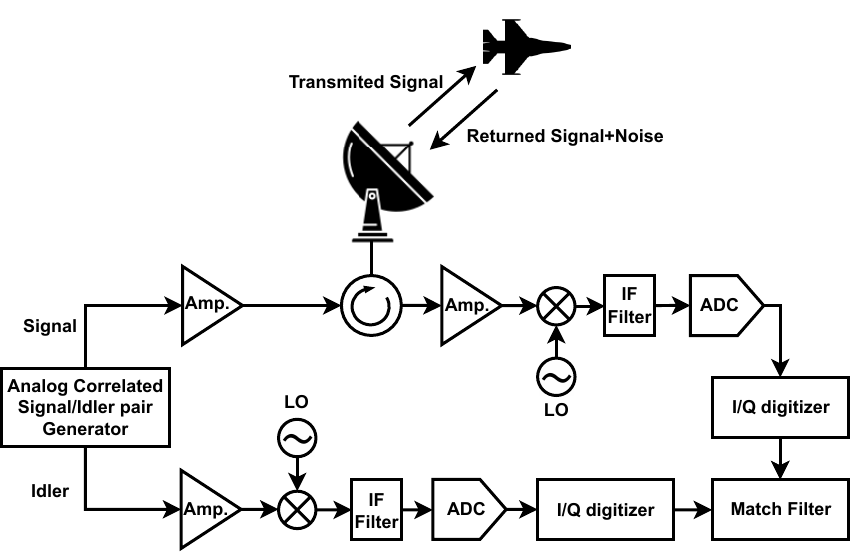}
	\centering
	\caption{Block diagram of a mono-static noise radar. In a classical noise radar, the analog correlated signal/idler pair generator is a coherent state source together with a splitter or an ideal power divider that generates two identical coherent states, while in a quantum noise radar, it is an entangled microwave photon-pair source.  }
	\label{fig3}
\end{figure*}

In direct detection radars, the returned signal can be easily masked by any noise source or unwanted reflected waveform into the receiver. Thanks to the correlation, in noise radars \cite{narayananNoiseRadarDesign,lukinNoiseRadarTechnology,noiseRadarOverview,efficientProcessingNoiseRadar2018}, one can extract and recognize the desired signal from noises by correlating the returned transmitted-signal with the reference signal, the so-called idler, at the receiver (for more information see Refs.~\cite{bistaticNoiseRadar2023,balaji2022PerformancePrediction}).
The simplified schematic of a noise radar is shown in Fig.~(\ref{fig3}). 
It is worthwhile to remember that the difference between a classical noise radar and an entangled quantum radar is that,  in classical noise radars, the signal and idler are classically correlated EM fields, while, in the latter, their quantum states are quantum entangled EM waveforms \cite{quantumilluminationSethLloyd2008,barzanjehQIlluminationOMS2015,balajiMythToReality2020}. A detailed discussion about the costs, advantages, and disadvantages of quantum entangled noise radars can be found in \cite{quantumRadarCost2020,defenceReportCanada,balajiSnapshot2020,quantumAdvantageInQRadarNature2023}.

Usually, the amount of the correlation between the two signals is quantified by the covariance matrix, $ \mathcal{C} $, formed by the in-phase and quadrature voltages of the signal and idler. If we denote the in-phase and quadrature voltages by $ I_j $ and $ Q_j $ with $ j=s $ for signal and $ j=i $ for idler, then the covariance matrix formed by these voltages is written as \cite{quantumEnhancedNoiseRadar2019,sorelli2022}
\begin{align} \label{covarianceMatrix1}
& \mathcal{C} = \left( {\begin{array}{*{20}{c}}
	C_{ss} & C_{si} \\
	C_{is} & C_{ii} 
	\end{array}} \right),
\end{align}
in which $C_{ij}$-s are $2\times2$ blocks given by
\begin{align} \label{covarianceMatrix1blocks}
& C_{ij} = {\begin{array}{*{20}{c}}
	\braket{I_i I_j} & \braket{I_i Q_j}  \\
	\braket{Q_i I_j} & \braket{Q_i Q_j} 
	\end{array}}.
\end{align}
The off-diagonal blocks of $\mathcal{C}$ (i.e., $C_{si}$ and $C_{is}$) represent the correlation between the signal and idler $I$-$Q$ voltages: $ \hat I=(\hat a +\hat a^\dagger)/\sqrt{2} $ and $ \hat Q= (\hat a - \hat a^\dagger)/\sqrt{2 i}$ in which $ \hat a $ ($ \hat a^\dagger $) is the microwave field annihilation (creation) operator. The amount of correlation between the signal and idler is determined by the well-known Pierson correlation coefficient, $ \rho $, which is given by
\begin{align}  \label{ro_1}
& \rho = \dfrac{\langle {{I_s}{I_i}} \rangle}{\sqrt{\langle {{I_s}{I_s}} \rangle \langle {{I_i}{I_i}} \rangle }}.
\end{align}
Physically, the Pierson coefficient is the coincidence between the signal and idler quadratures in a homodyne output detection scheme. Using mathematical algebraic calculations one can show that it is proportional to the coincidence between the intensity of the signal and idler fields.

In the problem of target detection in noise radars, $ \rho$ has great importance: it is kind of a detector function by which we can identify the presence of the target in the interrogation region \cite{dawood2021ReceiverOperating}. 
If there is a correlation between the received signal and the retained idler, i.e., $ \rho \neq 0 $, then we conclude that there is a target in the interrogation region, otherwise, if $ \rho=0 $, then there is no target there. In fact, the role that $ \rho $ plays in noise radars is similar to that of signal power in direct detection radars which is discussed in Sec.~\ref{sec2.DirectDetectionRadars}. Therefore, in analogy to the radar range equation of direct detection radar (\eqref{power1}) which gives the radar signal power at the receiver as a function of the target range $ R $, the radar range equation for noise radars should give the correlation coefficient as a function of target range $ R $. This equation was obtained by D. Luong \textit{et.al.} \cite{balaji2022PerformancePrediction,balajiCorrelationExperiment}, but they did not consider the effect of atmospheric attenuation in their studies. In the following, we derive a more general and explicit equation by considering the atmospheric attenuation.

To derive an explicit expression for correlation coefficient $ \rho $ in terms of the target range R, we have to first write $ \rho $ in terms of the signal power $ P $, and then use the direct detection radar range equation \eqref{power1}. Here, we use some results obtained in \cite{balaji2022PerformancePrediction}. Generally, the received signal and idler can be decomposed into two different parts: a perfectly correlated part consist of $N_0$ photons, and a completely uncorrelated part.
Consequently, one can write the power of the received signal and idler respectively as
\begin{align} 
P_{\mathrm{s}}^{\mathrm{rec}}&=\eta(R) P_{\mathrm{s}}^{\mathrm{corr.}} + P_{\mathrm{n,s}}, \label{p_rec_signal} \\
P_{\mathrm{i}}&=P_{\mathrm{i}}^{\mathrm{corr.}}+P_{\mathrm{n,i}} ,\label{signal_idler_power}
\end{align}
in which $ P_{\mathrm{k}}^{\mathrm{corr.}}=G_{\mathrm{amp, k}}N_0 h f_k B $ with $\mathrm{k}=\mathrm{s}, \mathrm{i}$ for signal and idler, respectively, where $G_{\mathrm{amp, k}}$ is the amplification at the transmitter, $f_{\mathrm{k}}$ is the frequency, and $B$ is the source bandwidth (for more details about the amplification mechanism and received photon rates at the receiver antenna see Appendix.~B). Moreover, $  P_{\mathrm{n,s}}  $ and $ P_{\mathrm{n,i}} $ are the power of added noise to the signal and idler, respectively, which are uncorrelated, and $ \eta(R) $ is the transfer function of the transmitter-target-receiver channel and given in \eqref{eta1}. 
Note that $  P_{\mathrm{n,s}}  $ and $ P_{\mathrm{n,i}} $ can be seen as a total added noises into the idler and received signal powers.
If the power of the environmental or background noise added to the signal is $ P_n $, then we can write
\begin{align}  
& P_{\mathrm{n,s}}=\eta(R) P_{\mathrm{n,i}} + P_n.  
\end{align}
By following the treatment given in \cite{balaji2022PerformancePrediction} and after some straightforward calculations we get to
\begin{align} 
&  \rho =  \dfrac{\rho_0}{\sqrt{1+\dfrac{P_n}{\eta_{(R)} P_i}}} , \label{ro_2}
\end{align}
where $ \rho_0 $ is the normalized correlation coefficient of the signal and idler at the transmitter and given by \cite{balaji2022PerformancePrediction}
\begin{align} \label{ro_0}
&  \rho_0= 1-\dfrac{P_{\mathrm{n,i}}}{P_i},
\end{align}
which is maximum when there is no noise. Now, using \eqref{eta1} one can rewrite \eqref{ro_2} as
\begin{align} 
&   \rho (R)=  \dfrac{\rho_0}{\sqrt{1+ \dfrac{1}{F(R)^2} \left(\dfrac{R}{R_c^{\mathrm{NR}}}\right)^4}}, \label{ro_R}
\end{align}
in which $ F(R) $ is the form factor describing the atmospheric attenuation of the signal and defined in Sec.~\ref{sec2.DirectDetectionRadars}, and
\begin{align} 
& R_c^{\mathrm{NR}} = \left( \dfrac{\sigma G A_e P_i}{16\pi^2 P_n}  \right)^{1/4} \label{r_c_NR}
\end{align}
is the characteristic range of the noise radar and is similar to the case of the direct detection radar (see \eqref{r_c_direct}), except that here $ P_i $ is the idler power. $ R_c^{\mathrm{NR}}  $ is the range at which the correlation between the signal and idler under the condition of in-vacuum propagation ($ \gamma=0 $) is reduced by a factor of $ \sqrt{2} $ with respect to their correlation at the transmitter, i.e., $ \rho(R_c^{\mathrm{NR}})=\rho_0/\sqrt{2} $ (for $ \gamma=0 $).

It is worth noting that \eqref{r_c_NR} is applicable for both CW classical noise radars and quantum entangled radars. The only difference between them is that quantum radars can achieve higher correlation coefficients $ \rho_0 $ compared to the classical ones. This difference becomes intuitive in the next subsection when we drive an explicit expression for the maximum resolvable range for noise radars.

\begin{figure*}
	\centering 
	\subfloat[]
	{\includegraphics[width=6.75cm]{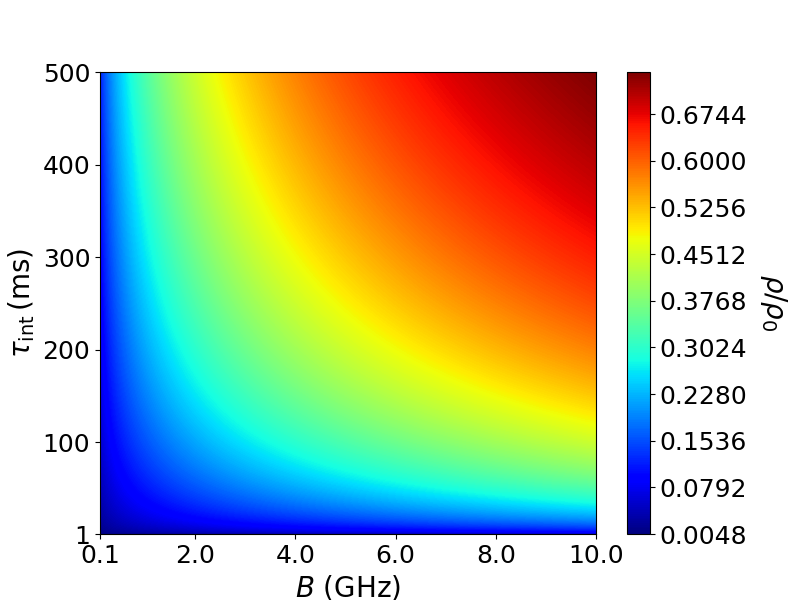}} 
	\subfloat[]
	{\includegraphics[width=5.5cm]{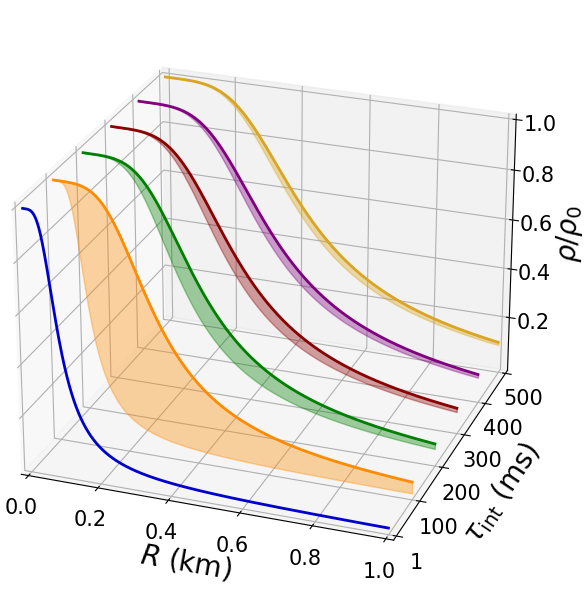}}   
	\subfloat[]
	{\includegraphics[width=5.5cm]{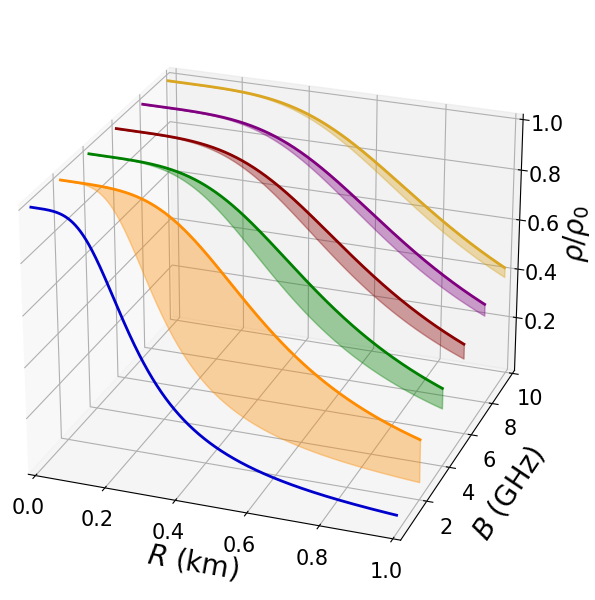}}
	\caption{(Color online) (a) Contour plot of the normalized correlation coefficient, $\rho/\rho_0$, vs detection integration time, $\tau_{\mathrm{det}}$, and signal/idler bandwidth, $B$, for the detection range of $R=1~\mathrm{km}$. (b) $\rho/\rho_0$ with respect to the target range $R$ for $B=100~\mathrm{MHz}$ and the different values of the detection integration time as $\tau_{\mathrm{int}}=\{1, 100, 200, 300, 400, 500 \} ~\mathrm{ms}$ respectively for blue, orange, green, brown, purple and yellow lines. (c) $\rho/\rho_0$ as a function of target range $R$ for $\tau_{\mathrm{int}}=100 ~\mathrm{ms}$ and different values of signal bandwidth $B$ as $B=\{0.1, 2,4,6,8,10 \} ~\mathrm{GHz}$ respectively for blue, orange, green, brown, purple and yellow lines. Other parameters of the system are the same as ones given in the table~\ref{table1}.  The shaded regions in (b) and (c) demonstrates the enhancement of correlation coefficient compared to its the previous curve. }
	\label{fig4}
\end{figure*}

\subsection{Detection Condition in Noise Radars}
Before discussing the detection condition in noise radars, one should consider the effect of detection time on the correlation coefficient. To this, using the definition of retained-idler SNR given in \cite{balaji2022PerformancePrediction} as $ {\mathrm{SNR_i}}= P_\mathrm{i}/P_\mathrm{n}$ , we rewrite the correlation coefficient in \eqref{ro_2} as
\begin{align} 
& \rho =  \dfrac{\rho_0}{\sqrt{1+\dfrac{1}{\eta(R) { \mathrm{SNR}_{\mathrm{i}}} }}}. \label{ro_rewrite}
\end{align}
If the target detection occurs by integrating over $ M= \tau_{\mathrm{int}} B_{\mathrm{det}} $ samples, then the effective idler SNR given by $ { \mathrm{SNR_i^{eff}}}=M.{\mathrm{SNR_i}} $. By substituting this into \eqref{ro_2}, the effective correlation coefficient is obtained as
\begin{align} 
& \rho_{\mathrm{eff}}(R)=\dfrac{\rho_0}{\sqrt{1+\dfrac{1}{\eta(R) {\mathrm{SNR_i^{eff}}}}}}.  \label{ro_eff_NR1}
\end{align}
Now, using the definition of $ \eta(R) $ and $\mathrm{{SNR}_i^{eff}}$, the effective correlation coefficient can be rewritten as
\begin{align} 
& \rho_{\mathrm{eff}} (R)=  \dfrac{\rho_0}{\sqrt{1+ \dfrac{1}{M F(R)^2} \left(\dfrac{R}{R_c^{\mathrm{NR}}}\right)^4}}. \label{ro_eff_NR2}
\end{align}

Figure \eqref{fig4} illustrates the effect of the integration time, $\tau_{\mathrm{int}}$, and signal bandwidth, $B$, on the normalized correlation coefficient, $\rho/\rho_0$. In Fig.~\ref{fig4}(a), we have plotted $\rho/\rho_0$ as a function of $\tau_{\mathrm{int}}$ and signal bandwidth $B$ for the range $R=1~\mathrm{km}$. As is seen, although for the low integration times ($\tau_{\mathrm{int}}\sim 1~\mathrm{ms}$) and narrow signal bandwidth ($B\sim100~\mathrm{MHz}$) the correlation coefficient is approximately vanishes, but by increasing $\tau_{\mathrm{int}}$ and broadening $B$ the correlation coefficient increases substantially such that for $\tau_{\mathrm{int}}=500~\mathrm{ms}$ and $B=10~\mathrm{GHz}$, which are experimentally feasible in practical applications, we get to $\rho\simeq0.7\rho_0$. Its reason is that the number of integration samples, $M$, and idler power, $P_i$, increase by growing $\tau_{\mathrm{int}}$ and $B$, respectively. 
To show the behavior of signal/idler correlation, we have plotted $\rho/\rho_0$ as a function of range, $R$, for the different values of $\tau_{\mathrm{int}}$ and $B$  in Figs.~\ref{fig4}(b) and (c), respectively. The shaded regions in these figures demonstrate the enhancement of correlation coefficient at the given $\tau_{\mathrm{int}}$ (in Fig.~\ref{fig4}(b)) or $B$ (in Fig.~\ref{fig4}(c)) compared to the previous $\tau_{\mathrm{int}}$ or $B$ considered in that plot.
For instance, the shaded orange region in Fig.~\ref{fig4}(b) illustrates the enhancement of $\rho/\rho_0$ at the integration time of $\tau_{\mathrm{int}}=100~\mathrm{ms}$ compared to the previous curve (blue) which is plotted for $\tau_{\mathrm{int}}=1~\mathrm{ms}$. The same interpretation is valid for other curves in both these figures.
Therefore, it is clear that the signal/idler correlation decreases by increasing the target range $R$. However, at a given range $R$, the correlation increases by increasing $\tau_{\mathrm{int}}$ and broadening $B$, which is in agreement with Fig.~\ref{fig4}(a). Therefore, one can conclude that the lossy effects due to the atmosphere absorption can be compensated by increasing the signal-bandwidth or detection time.

Note that in noise radars, the presence of a target is declared if the effective correlation coefficient ($\rho_{\mathrm{eff}} $) becomes greater than a threshold one, $ \rho_{\mathrm{th}}  $. This threshold depends on the number of integration samples $M$, and the false alarm probability, $ p_{\mathrm{fa}} $, and given by \cite{balaji2019EstimatingCorrelationCoefficients}
\begin{align} 
& \rho_{\mathrm{th}} =\sqrt{\dfrac{-\mathrm{ln}P_{\mathrm{fa}}}{M}}. \label{ro0_th}
\end{align}
Therefore, the detection condition in noise radars can be written as
\begin{align}
& \rho_{\mathrm{eff}}(R) \ge \rho_{\mathrm{th}}. \label{falseEq}
\end{align}  
Using \eqref{ro_eff_NR2} and after some algebraic calculations, the detection condition in classical noise radar is simplified as
\begin{align}  
& R \times 10^{\gamma R/20} \le M^{1/4} \left[ \left(\dfrac{\rho_0}{\rho_{\mathrm{th}}} \right)^2-1 \right]^{1/4} R_c^{\mathrm{NR}}. \label{rangeEq_1_inequalityNoiseRdara}
\end{align}
Surprisingly, by comparing the above inequality to \eqref{rangeEq_1} for the direct detection radar, we find that the quantity $ [ \left(\rho_0 / {\rho_{\mathrm{th}}} \right)^2-1 ]^{1/4} $ on the right-hand side of \eqref{rangeEq_1_inequalityNoiseRdara} exactly plays the role of the quantity $ 1/({\mathrm{{SNR}_{th}}})^{1/4} $ appeared on the right-hand side of \eqref{rangeEq_1}. This analogy motivates us to define a threshold SNR for target detection in a noise radar as
\begin{align} 
& \mathrm{{SNR}_{th}^{NR}}\equiv \left[  \left( \dfrac{\rho_0}{{\rho_{\mathrm{th}}}} \right)^2-1 \right]^{-1}.  \label{SNR_th_NR}
\end{align}
\begin{figure*}
	\centering 
	\subfloat[]
	{\includegraphics[width=8cm]{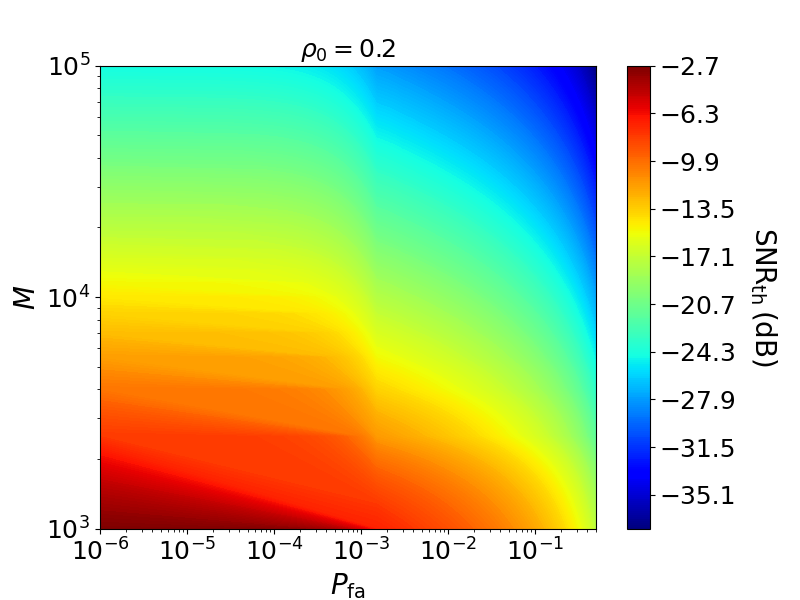}} 
	\subfloat[]
	{\includegraphics[width=8cm]{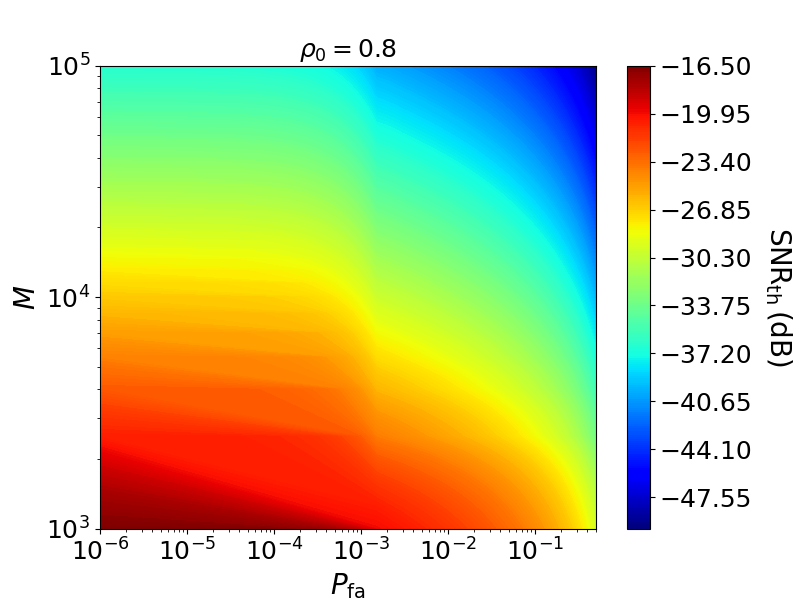}}   
	\caption{(Color online) The threshold-SNR of a noise radar versus the probability of false alarm $P_{\mathrm{fa}}$ (horizontal axis) and number of integration samples $M$ (vertical axis) (a) and (b), respectively for $\rho_0=0.2$ and $\rho_0=0.8$.}
	\label{fig55}
\end{figure*}
Fig~(\ref{fig55}) shows $\mathrm{{SNR}_{th}^{NR}}$ as a function of false alarm probability $P_{\mathrm{fa}}$ and number of integration samples $M$ for different values of the initial correlation coefficient of $\rho_0 = \{0.2,0.8\}$. 
Comparing figures \ref{fig55}(a) and (b) reveals that increasing the initial correlation coefficient at the source (i.e., $\rho_0$) significantly lowers the threshold $\mathrm{SNR}$ of the noise radar.
As is seen, the values of $\mathrm{SNR_{th}^{NR}}$ are much smaller than its counterpart, $\mathrm{{SNR}_{th}}$, in direct detection radars which belong to the domain $10~\mathrm{dB}$ to $20~\mathrm{dB}$ \cite{skolnikBook}. 
Now, depending on the value of the false alarm probability, we define two classes of noise radar: early alarm and track noise radar for which $P_{\mathrm{fa}}\sim 0.5 $ and $P_{ \mathrm{fa}}\ll0.5$, respectively. From Fig.~\ref{fig55}, it can be seen that the $\mathrm{SNR_th}$ for a early alarm noise radar with $M=10^5$ is $-35~\mathrm{dB}$ for $\rho_0=0.2$, while it decreases to $-47~\mathrm{dB}$ for $\rho_0=0.8$. For a track noise radar with $M=10^5$, the threshold $\mathrm{SNR}$ equals $-22~\mathrm{dB}$ and $-35~\mathrm{dB}$ for $\rho_0=0.2$ and $0.8$, respectively.  
It can be concluded that in a noise radar the threshold-SNR is improved about 4-6 order of magnitude compared to the direct radars which means that in a noise radar, thanks to the correlation, one can effectively identify a received signal with power much weaker compared to a direct radar that will yield the higher range.

By substituting \eqref{SNR_th_NR} into \eqref{rangeEq_1_inequalityNoiseRdara}, the detection condition in a noise radar becomes
\begin{align} 
& R \times 10^{\gamma R/20} \le \left(\dfrac{M}{{\mathrm{{SNR}_{th}^{NR}}}} \right)^{1/4} R_c^{\mathrm{NR}}.   \label{rangeEq_2_inequalityNoiseRdara}
\end{align}
It is surprising to see that the above inequality has the same form as \eqref{rangeEq_1} which is derived for direct detection radars. This analogy, together with the threshold-SNR for noise radars illustrated in Fig.~\ref{fig55} dictates that one can imagine of a noise radar as a direct detection radar with substantially enhanced threshold-SNR by replacing $\mathrm{{SNR}_{th}} \to \mathrm{{SNR}_{th}^{NR}}$ and $R_c^{\mathrm{dr}}\to R_c^{\mathrm{NR}}$.

\subsection{Maximum Detection Range in Noise Radars}
\begin{figure*}
	\centering
	\subfloat[]
	{\includegraphics[width=8cm]{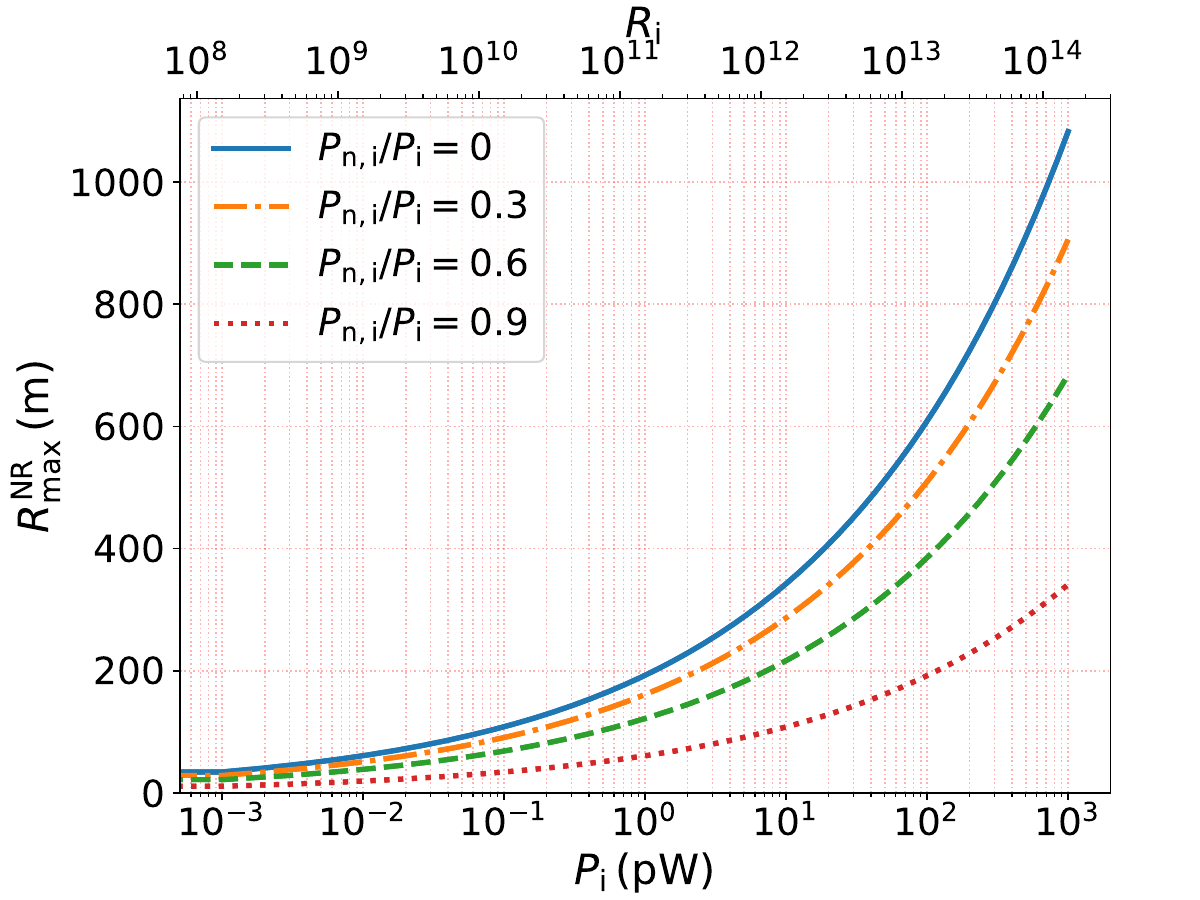}}
	\subfloat[]
	{\includegraphics[width=8cm]{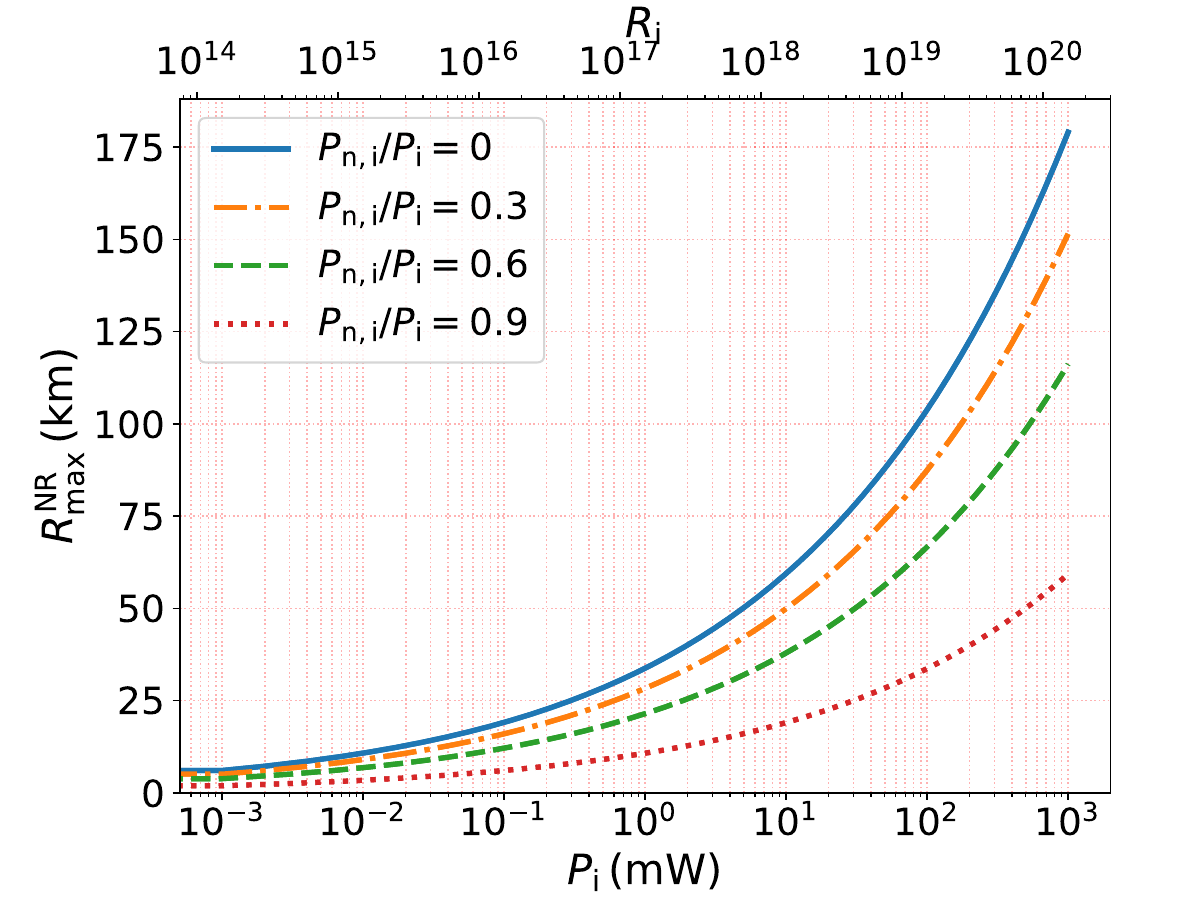}}
	\centering
	\caption{(Color online) Maximum range of a noise radar, $ R_{\mathrm{max}}^{\mathrm{NR}}  $, with respect to the idler power $ P_{\mathrm{i}} $ (lower horizontal axis) and the corresponding idler photon rate $R_{\mathrm{i}}$ (upper horizontal axis) in (a) low-brightness and (b) high-brightness regime for the different idler added noise powers as $P_{\mathrm{n,i}}/P_{\mathrm{i}}=\{0, 0.3, 0.6, 0.9 \}$ that respectively are referred to the blue-solid, orange dot-dashed, green-dashed and red-dotted lines. They are corresponds to the different initial correlation coefficient as $\rho_0=\{1, 0.7, 0.4, 0.1\}$, receptively. Here, the false alarm probability is set to the tracking case $P_{\mathrm{fa}}=10^{-6}$, and the other system parameters are the same as in the table.~\ref{table1}.}
	\label{fig5}
\end{figure*}

The lower bound of the inequality that given in \eqref{rangeEq_2_inequalityNoiseRdara} obtained for the maximum range $ R_{\mathrm{max}} $ at which the noise radar is able to identify the target. By solving this equation analytically, an explicit expression for the maximum resolvable range of a noise radar obtain as
\begin{align} 
&  R_{\mathrm{max}}^{\mathrm{NR}} = \dfrac{1}{0.115 \gamma} {\mathrm{ln}}\left[  1+ 0.115 \gamma \left( \dfrac{M}{{\mathrm{{SNR}_{th}^{NR}}}} \right)^{1/4} R_c^{\mathrm{NR}}  \right].   \label{rangeEqNR_final}
\end{align}
This equation is equivalent to the expression obtained for the maximum range of a direct detection radar (see \eqref{rangeEqDirectFinal}).

Figure \ref{fig5} shows the maximum range of a noise radar $R_{\mathrm{max}}^{\mathrm{NR}}$ with respect to the idler power $P_{\mathrm{i}}$ (lower horizontal axis) and the idler photon rate $R_{\mathrm{i}}\equiv P_{\mathrm{i}}/hf$ (upper horizontal axis) for different values of the power of the uncorrelated part of idler $P_{\mathrm{n,i}}$  in the both low- and high-brightness regimes. 
As is evidence, both increasing the idler power $P_{\mathrm{i}}$ and suppressing the idler added noise power $P_{\mathrm{n,i}}$, which is equivalent to increasing the $\rho_0$, can enhance the maximum detection range of noise radars. By comparing this figure with Fig.~\ref{fig2}, it is revealed that the maximum detection range of a noise radar is significantly higher than a direct detection radar with the same parameters in both low- and high-brightness regimes.
Significantly, in the low-brightness regime, although the maximum detection range of the considered direct detection radar is limited to $10~\mathrm{m}$ (at the best case), a noise radar with the same features can recognize targets up to a range of $1000~\mathrm{m}$. 
On the other hand, while the maximum detection range of a direct detection radar in the high-brightness regime is limited to $1.5~\mathrm{km}$ (at the best case), it rises to $175~ \mathrm{km}$ for a noise radar with the same features in Fig.~\ref{fig2}.
This enhancement originates from the substantial decrease of the threshold signal-to-noise ratio in noise radars.
It is reminded that lowering the threshold $\mathrm{SNR}$ significantly enhances the maximum range of a direct detection radar (see Fig.~\ref{fig2}) and, as is mentioned a noise radar behaves effectively as a direct detection radar with enhanced $\mathrm{{SNR}_{th}}$. That is why a noise radar can achieve much longer ranges with respect to the conventional radars.
Moreover, Fig.~(\ref{fig5}) shows that the maximum range of a noise radar increases if the $initial$ correlation between the signal and idler generated by the source at the transmitter grows substantially. One way to generate signal and idler pairs with enhanced correlation coefficient is employing microwave entangled sources, which is the idea of quantum entangled radar \cite{wilsonMicrowaveQuantumRadar2018,barzanjehQIlluminationOMS2015,frasca2020EntangledCoherentQRadar}. 
We will clarify this issue in the next section by deriving an explicit expression for the correlation coefficient generated by QTMS radar and comparing the obtained results with classical noise radar.

\section{QTMS Radar Vs. Classical Noise Radar}  \label{sec4.QuantumNoiseRadar}
Consider a CW QTMS radar that exploits a two-mode squeezed vacuum (TMSV) state of electromagnetic field, i.e., an entangled signal-idler photons pair (the squeezing origination is discussed in Appendix.~A). 
Quantum description of this state in the number basis is given by \cite{sorelli2022,shySPDC}
\begin{align} 
& \vert \psi_Q \rangle = \sum_{n=1}^{\infty} \sqrt{\dfrac{N_s^n}{(1+N_s)^{n+1}}}  \vert n_s \rangle \vert  n_i \rangle ,   \label{SPDSstate}
\end{align}
in which $ \vert n_s \rangle  $ and $ \vert n_i \rangle  $ represents the number state of the signal and idler modes, respectively, and $ N_s $ gives the mean number of photons per mode. Here, the \textit{mode} refers to the signal and idler subspace modes. $N_s$ can be easily obtained from the field power $P$ via $N_s = P / (hfB) = R/B$, in which $B$ is the signal bandwidth and $R=P/(hf)$ is the photon rate. In other words, photon per mode $ N_s $ means that if we define the mean photon number in each mode as $ \langle \hat n_i \rangle:=N_i$ and $ \langle \hat n_s \rangle=:N_s $, then it can be easily obtained that $ N_s=N_i $. 
Here, it should be noted that the ``mode" in the QTMS state, which refers to the one of degrees of freedom in the subspace of quantum states of light wave, i.e., the signal or idler subspaces, can be labeled by spatial mode of light wave that refers to the spatial distribution of light intensity at a given frequency $ \omega $. A spatial mode can be defined in different spaces such as a waveguide and an optical cavity, and also can be divided into two types of temporal and spatial modes. If two spatial modes have the same spatial distribution but different frequencies, then they are called temporal modes, while for the same frequency but different spatial distributions, they are called spatial modes.
Generally, one can show that the number of modes in a quantized free space with volume $ V=L^3 $ between frequency $ \omega $ and $ \omega + \Delta\omega $ can be defined as $ \mathcal{M}=2 (L/2\pi)^3 d^3k $. By assuming $ L=c\Delta t $, $ \Delta t \simeq  \Delta \omega^{-1} $, and $ d^3k=k^2 dk d\Omega_k $ in the regime of $ \lambda \ll L $ we get to $  \pi \sqrt{\mathcal{M}}=(\omega/\Delta \omega)=(\lambda/\Delta \lambda) $. 
For example, in a SPDC in optical domain of the wavelength of $ \lambda\simeq 810 $nm with $ \Delta \lambda \simeq 70-80 $nm or 10nm the number of spatial modes is about 13 or 600 respectively, while for the microwave frequency $ \omega=10 $GHz with $ \Delta \omega=B=1 $GHz or 100MHz there exists 10 or 1014 spatial modes, respectively. It is clear that larger bandwidth results in a fewer spatial modes toward single mode approximation for each signal/idler mode.

Here, it should be noted that such an entangled state can be easily generated in the optical domain using the well-known nonlinear optical process of spontaneous parametric down-conversion (SPDC) \cite{aliBBO,aliPPKTP}. 
%
%
In contrast to the other sources such as superconductors \cite{barzanjehExperimentQuantumRadar2020,wilsonMicrowaveQuantumRadar2018} or electro-optomechanical systems \cite{barzanjehQIlluminationOMS2015} that are used for entanglement generation in microwave domain, SPDC sources are more inexpensive, user-friendly, and notably work at room temperature (i.e., no need for ultra-low temperature cooling), and finally could be implemented on-chip via integrated photonics techniques and interestingly can be commercialized.
During these three decades, the SPDC in NLCs has been a variety range of applications in quantum metrology and measurement \cite{quantumellipsometry2018,aliDNA,spectroscopy5,imaging1}, quantum communication \cite{crypto9,zbindencrypto,villoresi1,motazediFSQKD} and so on (for more details and a review, see Ref.~\cite{quantummetrologybook}). 
However, beyond these platforms, recently it has been experimentally shown that NV-centers which work at room temperature can be used for radio-detection and ranging \cite{radarNV1_2023} and may open a new perspective for a new type of quantum radars based on NV-centers \cite{aliNVradar} (also see the references \cite{rfSensingNV2022,rfSensingNV2023,nVmagnetometry1,NVreceiverfT2024,NVreceiverPT2024}).

\begin{figure*}
	\includegraphics[width=8cm]{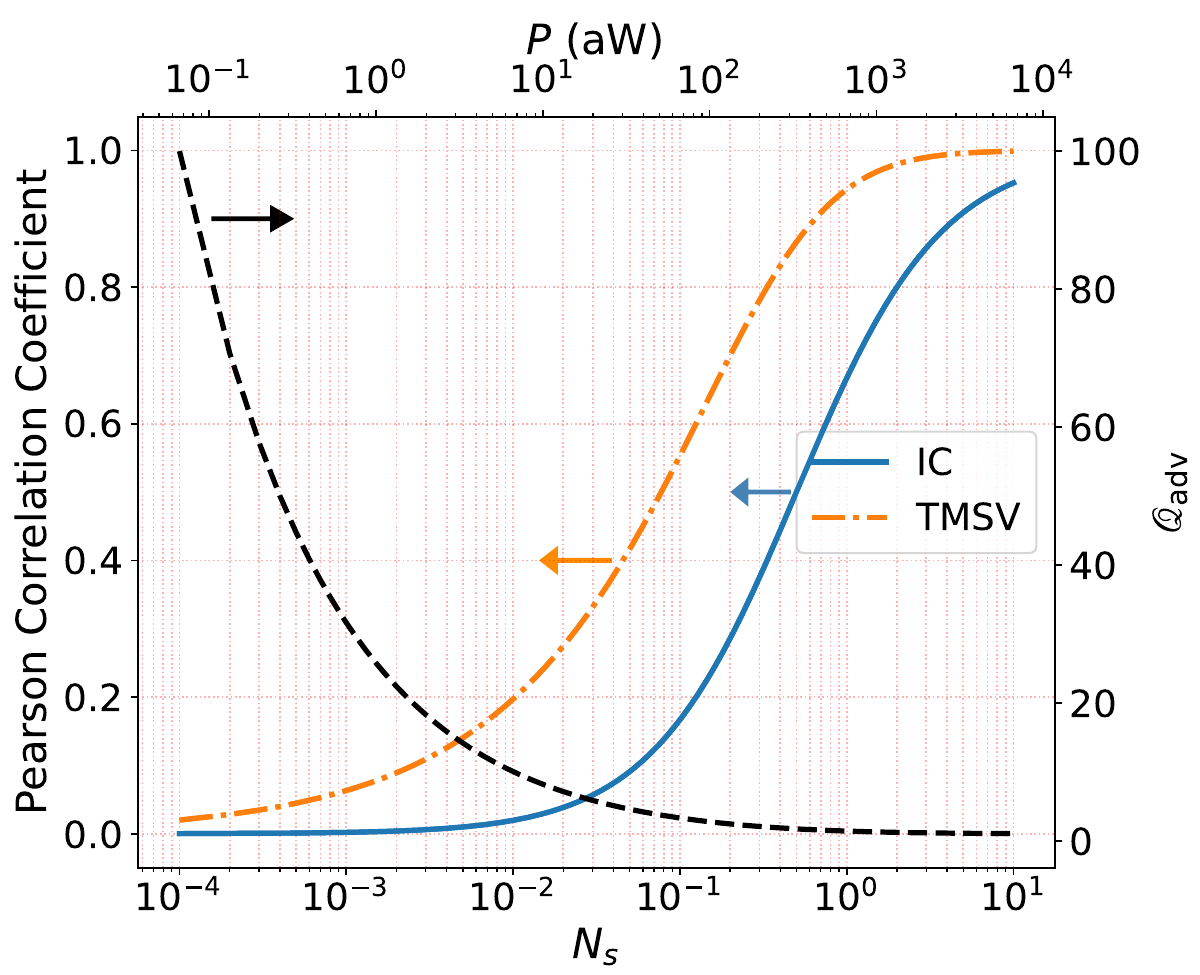}
	\centering
	\caption{(Color online) Pearson correlation coefficient (left vertical axis) for TMSV (\eqref{ro_TMSV}) and classical IC states  (\eqref{ro_c}) and quantum advantage $\mathcal{Q_{\mathrm{adv}}}$ (right vertical axis), with respect to the photon number per mode $N_s$ (lower horizontal axis) and the corresponding power $P=N_shfB$ (upper horizontal axis). Here, we consider the signal frequency and the signal bandwidth as $f=10~\mathrm{GHz}$ and $B=100~\mathrm{MHz}$, respectively.}
	\label{fig6}
\end{figure*}

Let us back to QI radar range equation formalism. According to the definition of the Pearson correlation coefficient $ \rho $ given in \eqref{ro_1}, for the TMSV state one can easily show that \cite{sorelli2022,qAdvantageMotazedi}
\begin{align} 
& \rho_{\mathrm{TMSV}}=\dfrac{2 \sqrt{N_s (N_s +1)}}{2N_s +1}. \label{ro_TMSV}
\end{align}
For an apple-to-apple comparison, we have to obtain the Pearson correlation coefficient for the two identical coherent states. Such states can be created by equally splitting an initial coherent state. One can show that the Pearson correlation coefficient for these classically identical coherent states (IC), i.e., classical illumination (CI), is given by \cite{sorelli2022}
\begin{align}
& \rho_{\mathrm{CI}}=\dfrac{2 N_s}{2N_s +1},     \label{ro_c}
\end{align}
in which $ N_s $ is again the mean photon number per each coherent state mode. Comparing \eqref{ro_TMSV} and \eqref{ro_c} reveals that the signal and idler prepared in an entangled TMSV state shows stronger correlation with respect to those prepared in  classical IC states, since $\rho_{\mathrm{TMSV}} > \rho_{\mathrm{CI}} $.

To understand the superiority of correlation between the signal-idler in the TMSV state over the classical IC states, it is useful to define the \textit{quantum advantage} $ \mathcal{Q_{\mathrm{adv}}} $ as
\begin{align} 
& \mathcal{Q_{\mathrm{adv}}}  \equiv \dfrac{ \rho_{\mathrm{TMSV}}}{\rho_{\mathrm{CI}}}=\sqrt{1+ \dfrac{1}{N_s}}.   \label{q_advantage1}
\end{align}

In Fig.~(\ref{fig6}), we illustrate the correlation coefficient $\rho$ for TMSV (orange dot-dashed line) and classical IC states (blue solid line), as well as the quantum advantage $\mathcal{Q_{\mathrm{adv}}}$ (black dashed line) with respect to the photon per mode $N_s$ and corresponding power $P=N_s hfB$. In this figure, we considered the signal frequency and bandwidth as $f=10~\mathrm{GHz}$ and $B=100~\mathrm{MHz}$, respectively. 
This figure reveals that the correlation coefficient for both QTMSV and classical IC states increases by increasing the photon number per mode, $N_s$ (or equivalently, the signal power $P$), and ultimately approaches the unity for sufficiently large $N_s$. 
Also, as is evident from Quantum advantage and the Pearson correlation coefficients of $ \rho_{\mathrm{TMSV}} $ and $ \rho_{\mathrm{CI}} $ in all given photon per mode, QTMSV states always surpasses the classical coherent one notably for small photon per mode, i.e., low-brightness regime, in which the QI can be better than the CI. While for large values of signal power, the quantum TMSV states show the same correlations as classical IC states, i.e., the advantage of quantum TMSV states over classical IC states vanishes, and thus QI has no advantage over CI. This implies careful engineering of the system parameters to achieve the quantum advantage of QI radar over CI radar to exploit the best quantum performance of entanglement, however, this engineering and optimizing seems to be a hard task.

%


In conclusion up to now, one can say that the signal-idler waveforms prepared in QTMSV states show stronger correlations compared to ones prepared in classically IC states. 
On the other hand, we have shown in Fig.~\ref{fig5} that in noise radars, the maximum detection range increases if the correlation between the signal and idler in the source becomes stronger. 
Consequently, we expect that the maximum detection range of a QTMS radar in which the signal-idler prepared in a TMSV state to be greater than an ideal classical noise (CN) radar in which the correlated states are IC states. 
For more intuition, let us write an explicit expression for the maximum detection range of a QTMS radar and an ideal CN radar separately. 
Using \eqref{rangeEqNR_final}, \eqref{ro_TMSV} and \eqref{ro_c}, one can write
\begin{align}
R_{\mathrm{max}}^{\mathrm{a}} &= \dfrac{1}{0.115 \gamma} {\mathrm{ln}} \left[ 1+0.115 \gamma \left( \dfrac{M}{\mathrm{{SNR}_{th}^{a}}} \right)^{1/4} R_c^{\mathrm{NR}} \right] , \label{R_max} \\
\mathrm{{SNR}_{th}^{a}}&=\left[  \left( \dfrac{\rho_{\mathrm{a}}}{\rho_{\mathrm{th}}} \right)^2 - 1  \right]^{-1}  \label{SNR_TMSV&Cl},
\end{align}
in which $ \mathrm{a= TMSV, CI} $, and  $R_c^{\mathrm{NR}}=[\sigma G A_e P_i/(16\pi^2P_n)]^{1/4}$ is the characteristic range. 
Note that the idler power depends on the generated mean number of photons per mode, $N_s$, through $P_i=G_{\mathrm{amp}}N_s h f_i B$ with $G_{\mathrm{amp}}$ being the total amplification gain between the photon pair in source and the radar antenna, and $B$ as before is the signal (or idler) bandwidth where usually are the same.
In QTMS radars implemented up to date, this gain is reported in the range of $40-80~\mathrm{dB}$ \cite{barzanjehExperimentQuantumRadar2020, balajiMythToReality2020}.

\begin{figure*}
	\centering
	\subfloat[] 
	{\includegraphics[width=8cm]{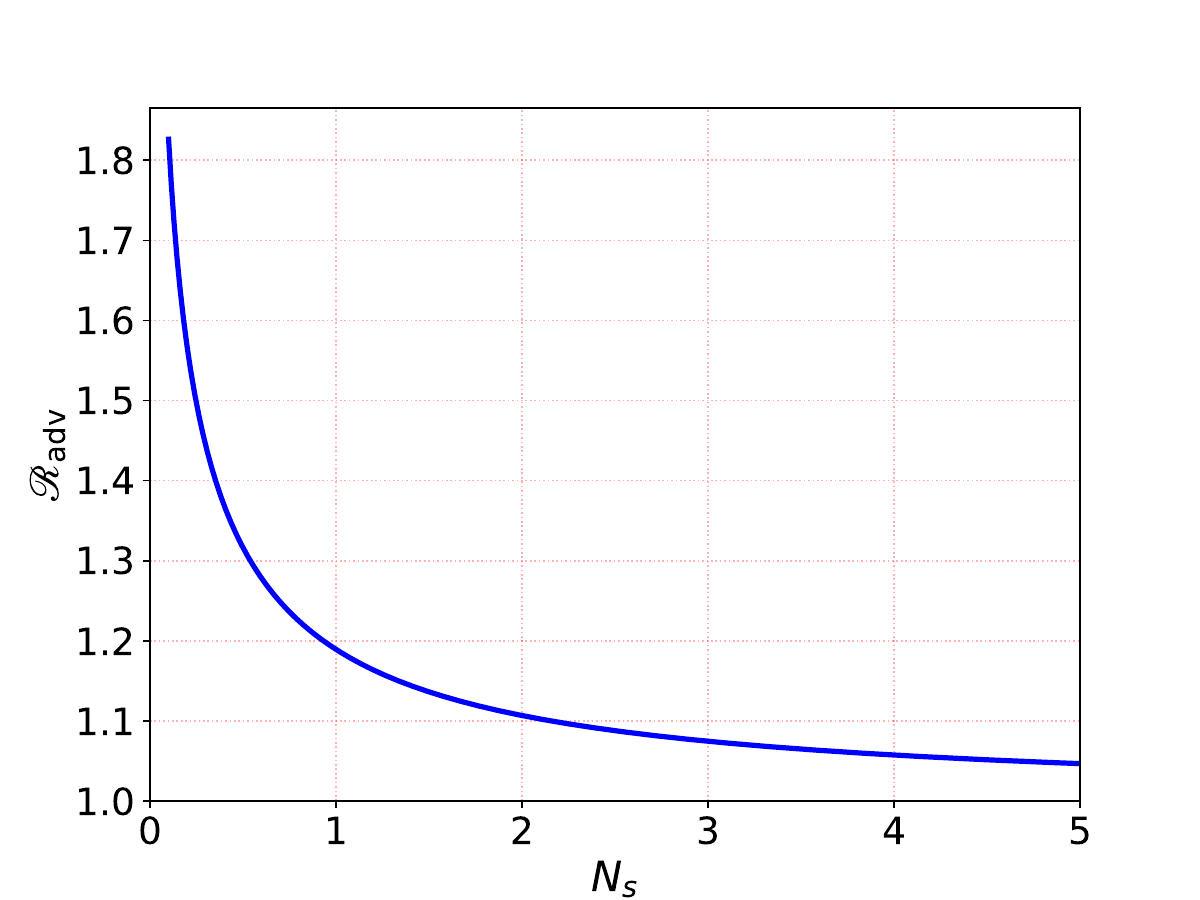}}\quad
	\subfloat[]
	{\includegraphics[width=8cm]{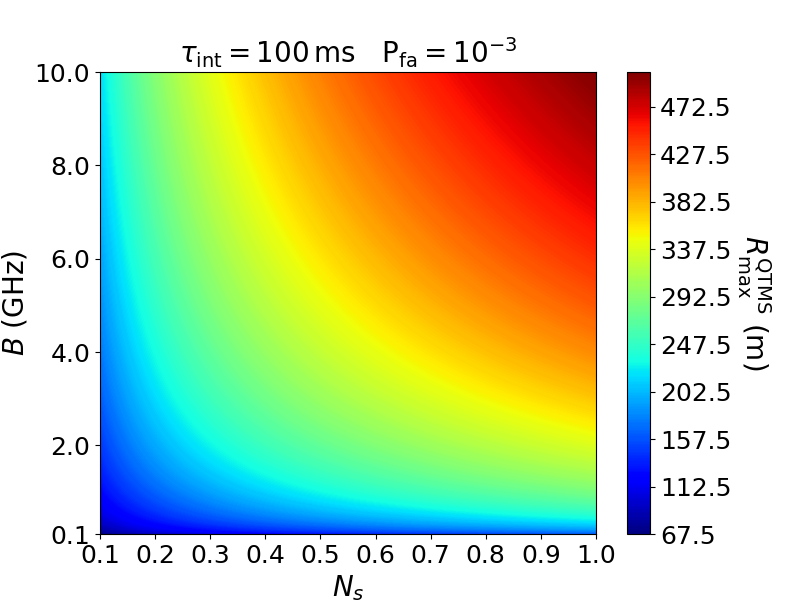}}\par\medskip
	\subfloat[]
	{\includegraphics[width=8cm]{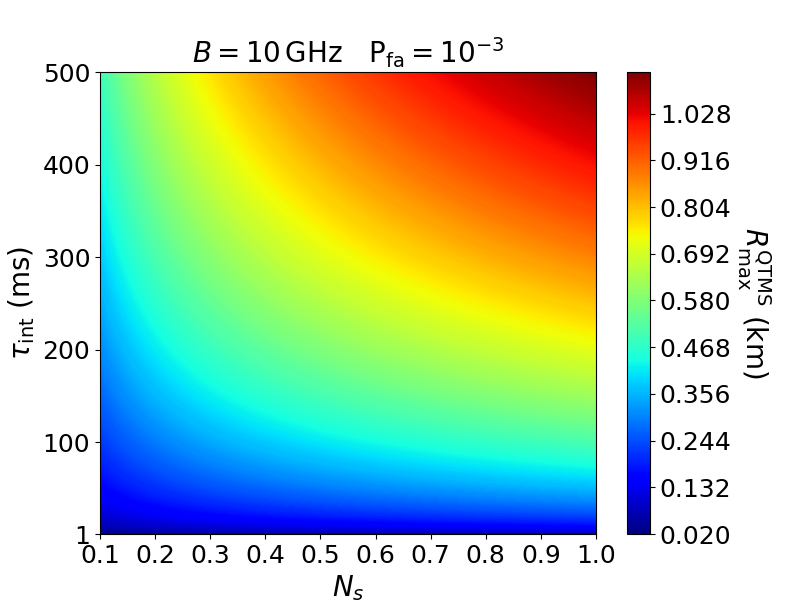}}\quad
	\subfloat[]
	{\includegraphics[width=8cm]{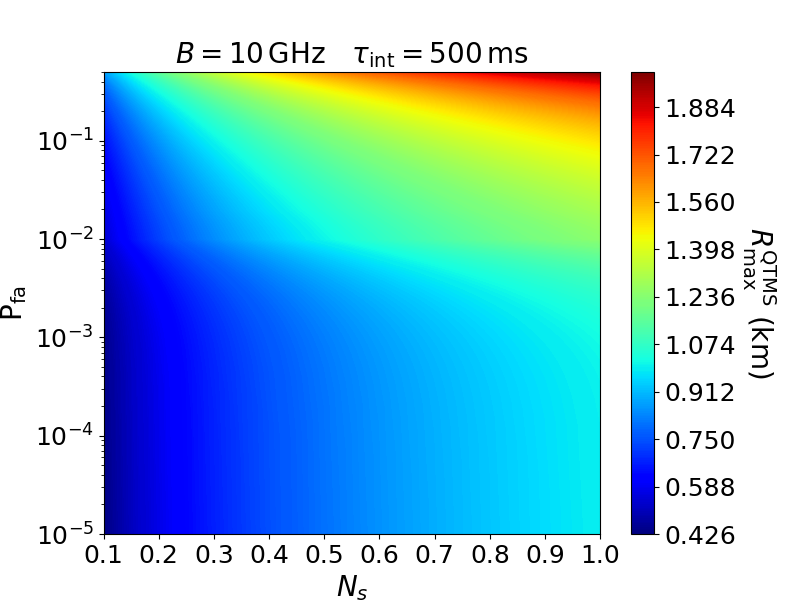}}
	\caption{(Color online) (a) Range advantage $\mathcal{R}_{\mathrm{adv}}$ as a function of generated photon number per mode. (b)-(e) Evaluation of the effect of different controllable system parameters, i.e., $B$, $N_s$, $\tau_{\mathrm{int}}$, and $P_{\mathrm{fa}}$, on the maximum detection range $R_{\mathrm{max}}$ of the QTMS radar system. Other parameters of the system is given in table \ref{table1}. }
	\label{fig7}
\end{figure*}

Now, it is advantageous to define a parameter that describes the superiority of the maximum detection range of a QTMS radar over an ideal CN radar with the same features, i.e., the Quantum-Range advantage as
\begin{align}  
& \mathcal{R}_{\mathrm{adv}}\equiv \frac{R_{\mathrm{max}}^{\mathrm{QTMS}}}{R_{\mathrm{max}}^{\mathrm{NR}}}.   \label{Range_advantage}
\end{align}

Fig.~\ref{fig7}(a) demonstrates the quantum range advantage $\mathcal{R}_{\mathrm{adv}}$ as a function of generated mean photon per mode $N_s$ for $N_s>0.1$, which reveals that $\mathcal{R}_{\mathrm{adv}}$ is significant in the low-brightness regime (i.e., at smaller $N_s$), and decreases gradually by increasing $N_s$ toward the high-brightness regime. 
For instance, the maximum detection range of a QTMS radar is about $1.8$ times greater than a CN radar when $N_s=0.1$, while $\mathcal{R}_{\mathrm{adv}}$ is lower than $1.2$ for $N_s>1$. That is why in QTMS radars it is vital to keep $N_s$ low via source engineering which will be discussed later. 
Note that $ N_s $ had been around 0.33  in the first experimental demonstration of QTMS radar using the Josephson superconductor-based technologies \cite{wilsonMicrowaveQuantumRadar2018}. 

From \eqref{R_max}, on one hand, it is clear that the maximum range of both QTMS and CN radars increases by increasing $R_c^{\mathrm{NR}}$ through increasing $N_s$. 
While, on the other hand, increasing the photon per mode $N_s$, leads to diminish of the quantum-range advantage. 
It means that there is a competition between maximum detection range and range advantage. This dictates the necessity of careful engineering of the quantum entangled source parameters and the other system parameters to have simultaneous quantum advantage and substantial detection range for practical applications out of the lab in the real world. 
Therefore, the quantity of `photon per mode' or the bandwidth of the generated entangled signal-idler RF-waveform in the source play the key role in quantum entangled radar to have advantage performance over the classical radar. 
Note that the photon per mode in the source $N_s$ relates to the parameters of the signal-photon rate and the signal bandwidth via $N_s=R_s/B$. Usually, during an experiment, the photon rate is an adjustable parameter up to its maximum value, 
while the signal bandwidth is a fixed parameter that is engineered before the source fabrication.
Then, by noting that the photon rate can be set to its maximum value, thus, the only way to increase $R_c^{\mathrm{NR}}$ is engineering and fabricating a entangled microwave source with the increased signal bandwidth $B$ which directly is equivalent to the decrease of the photon per mode. That is why having a source with improved bandwidth plays a gold role in a quantum entangled radar.

 As demonstrated in \cite{zorin2019flux}, it is experimentally feasible to fabricate a JTWPA with a 3dB-bandwidth of $B_{\mathrm{3dB}} \approx0.5f_p$ with center frequency of $f_s=0.5f_p$, where $f_p$ is the JTWPA pumping frequency. Therefore, for example, a JTWPA in degenerate regime with pump frequency as $f_p=20~\mathrm{GHz}$ can be considered as a source for generating entangled signal-idler photon-pairs with signal-idler frequencies $f_s=f_i=10~\mathrm{GHz}$ with the wide-bandwidth of $B=10~\mathrm{GHz}$. Also, some novel demonstrations of JTPWAs with ultra-wide bandwidth can be found in \cite{livreri2023JTWPA, zorin2019flux,zorin2016josephson}.

In figures \ref{fig7}(b)-(d), we have investigated the effect of different controllable parameters of a QTMS radar system (i.e., signal bandwidth $B$, integration time $\tau_{ \mathrm{int}}$, generated photon number per mode $N_s$, and probability of false alarm in target detection $ P_{\mathrm{fa}}$) on its maximum detection range $R_{\mathrm{max}}^{\mathrm{QTMS}}$. 
From these figures, one can see that $R_{\mathrm{max}}^{\mathrm{QTMS}}$ increases by rising $N_s$, which is in agreement with the previous paragraph. Figure \ref{fig7}(b) demonstrates that $R_{\mathrm{max}}^{\mathrm{QTMS}}$ reaches to $480~\mathrm{m}$ for $B=10~\mathrm{GHz}$, $\tau_{\mathrm{int}}=100~\mathrm{ms}$, $P_{\mathrm{fa}}=10^{-3}$ (an early alarm radar), and $N_s=1$ (for which $\mathcal{R_{\mathrm{adv}}}=1.2$).
Figure \ref{fig7}(c) illustrates that when an engineered entangled microwave photon-pair source as one demonstrated in \cite{livreri2023JTWPA} with $B=10~\mathrm{GHz}$ is employed, it is possible to reach $R_{\mathrm{max}}^{\mathrm{QTMS}}\sim 1~\mathrm{km}$ even for $N_s=0.67$ (for which $\mathcal{R_{\mathrm{adv}}}=1.25$) if the integration time is as long as $\tau_{\mathrm{int}}=500~\mathrm{ms}$.
According to Figure \ref{fig7}(d), in a QTMS radar with ultimate parameters of $B=10~\mathrm{GHz}$ and $\tau_{\mathrm{int}}=500~\mathrm{ms}$, if the probability of false alarm is considered to be $P_{\mathrm{fa}}=10^{-5}$ (tracking QTMS radar), the system is able to detect targets with maximum range of about $990~\mathrm{m}$ when $N_s=1$ ($\mathcal{R_{\mathrm{adv}}}=1.2$). However, for $ P_{\mathrm{fa}}=0.5$ (early alarm QTMS radar), the radar system is able to identify targets at the maximum range of $2~\mathrm{km}$ for $N_s=1$ ($\mathcal{R_{\mathrm{adv}}}=1.2$) and $863~\mathrm{m}$ for $N_s=0.1$ ($\mathcal{R_{\mathrm{adv}}}=1.8$).

As is seen, the bandwidth $B$ of the microwave photon-pair source employed in a QTMS radar, the integration time $\tau_{\mathrm{int}}$ assigned to the target detection, and probability of false alarm $P_{\mathrm{fa}}$ are three critical parameters that determine the ultimate detection range of the system. 
Based on the value of $P_{\mathrm{fa}}$, we classified QTMS radars into the early-alarm or search QTMS radar ($P_{\mathrm{fa}}\sim0.5$) and track QTMS radar ($P_{\mathrm{fa}} \ll 1$). We demonstrated that an early-alarm QTMS radar with extreme parameters of $B=10~\mathrm{GHz}$, $\tau_{\mathrm{int}}=500~\mathrm{ms}$, and other features considered in Fig.~(\ref{fig7}) is able to reveal targets (with $\sigma \simeq 0.5~\mathrm{m}^2$) with maximum range of $2~\mathrm{km}$ and $863~\mathrm{m}$ for $N_s=1$ ($\mathcal{R_{\mathrm{adv}}}=1.2$) and $N_s=0.1$ ($\mathcal{R_{\mathrm{adv}}}=1.8$), respectively.
This system is very desirable for detecting unmanned aerial vehicles (UAVs) such as quadcoupters in urban areas. Moreover, the mentioned system is able to track ($P_{\mathrm{fa}}=10^{-5}$) such targets with maximum range of $992~\mathrm{m}$ and $427~\mathrm{m}$ for $N_s=1$ and $N_s=0.1$, respectively. These imply that the long-range quantum entangled noise radar is achievable with the current technology for practical applications which will be discussed with more details in the next section.

\begin{table*}
	\caption{Maximum detection range estimation for some dominant experimentally realized QTMS radar systems. JPC: Josephson Parametric Converter; JPA: Josephson Parametric Amplifier; JTWPA: Josephson Traveling Wave Parametric Amplifier. The last column is our proposed system based-one the current technologies. Here, we assume that RCS is $ \sigma=0.1~\mathrm{m}^2 $ corresponding to a small UAV, and the fulse alarm probability of $P_\mathrm{fa}=10^{-3}$.}
	\label{table3}
	\setlength{\tabcolsep}{3pt}
	\begin{tabular}{|p{145pt}|p{30pt}|p{30pt}|p{50pt}|p{50pt}|p{50pt}|p{70pt}|}
		\hline
		Parameter& 
		Notation& 
		Unit & 
		Barzanjeh \textit{et.al.} \cite{barzanjehExperimentQuantumRadar2020} &
		Luong \textit{et.al.} \cite{balaji2019receiver} &
		Livreri \textit{et.al.} \cite{livreri2023JTWPA} &
		Our proposed \par system parameters \\
		\hline
		\hline
		Microwave photon pair source  & - & - & JPC & JPA & JTWPA & JTWPA \cite{liveri202210GHz}\\
		Signal frequency & $f_s$ & GHz & 10.09 & 7.5376 & 3.3 & 9  \\ 
		Idler frequency & $f_i$ & GHz & 6.8 & 6.1445 & 3.45 & 9  \\
		Signal bandwidth & $B$ & MHz & 20 & 1 & 3,000 & 9,000 \cite{zorin2019flux} \\ 
		Signal/idler amplification gain @ transmitter & $G_{\mathrm{amp}}$ & dB & 77.16 & 63 & 30 & 77 \\
		Environmental noise mean photon number & $N_b$ & - & 672 & 1015 & 100 & 693 \\ 
		Integration time & $\tau_{\mathrm{int}}$ & ms & 1,900 & 50 & - & 500 \\
		Detection bandwidth & $B_{\mathrm{det}}$ & kHz & 200 & 1,000 & - & 1,000 \\ 
		Number of integration sample & $M$ & - & $3.8\times10^5$ & $5\times10^4$ &  $1.5\times10^7$ & $5\times10^5$ \\
		Generated photon number per mode & $N_s$ & - & 0.5 & 0.57 & 0.05 & 0.05 \\
		Quantum advantage (from \eqref{q_advantage1}) & $\mathcal{Q_{\mathrm{adv}}}$ & - & 1.73 & 1.65 & 4.58 & 4.58 \\
		Range advantage (from \eqref{Range_advantage}) & $\mathcal{R_{\mathrm{adv}}}$ & - & 1.32 & 1.29 & 2.14 & 2.14 \\
		Maximum detection range (from \eqref{R_max}) & $R_{\mathrm{max}}$ & m & 1039.7 & 50.2 & 641.8 & 1421.4 \\
		\hline
	\end{tabular}
\end{table*}


\section{Experimental Discussion on the feasibility of the long-range quantum entangled radar} \label{sec5.experiment}

\subsection{Range-comparison between the implemented current quantum entangled radars}

\begin{figure*}
	\includegraphics[width=8cm]{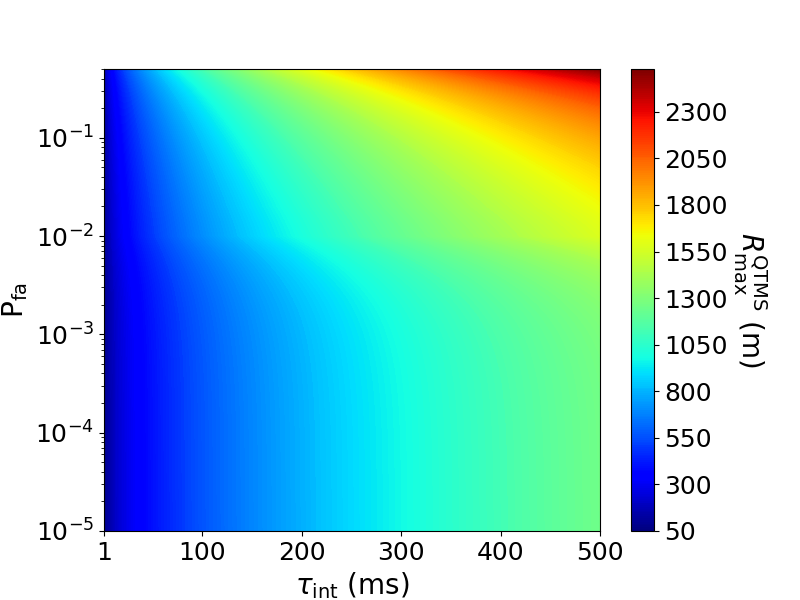}
	\centering
	\caption{(Color online) Maximum detection range of our proposed QTMS radar as a function of integration time $\tau_{\mathrm{int}}$ and probability of false alarm $ P_{\mathrm{fa}}$. Parameters of the system is given in table \ref{table3} (last column).}
	\label{fig8}
\end{figure*}

In this section, we are going to apply the formulation developed in Sec.~\ref{sec4.QuantumNoiseRadar} to estimate the maximum detection range of the three outstanding experimental demonstrations of QTMS radars \cite{barzanjehExperimentQuantumRadar2020, balaji2019receiver,livreri2023JTWPA}. 
The main features of these systems are summarized in table.~\ref{table3}. 
To evaluate the maximum range, we supposed that a Tx antenna with gain $G=15~\mathrm{dB}$, aperture efficiency $\epsilon_a=0.1$, and radius $r=0.1~\mathrm{m}$ is utilized to detect a small `UAV' with RCS of $\sigma=0.1~\mathrm{m}^2$ \cite{ryapolov2014radar}. Note that here we have considered the small RCS to show that even in this hard case, a quantum entangled radar can be applied for long-range practical applications.
Furthermore, we assume that the false alarm probability is $P_{\mathrm{fa}}=10^{-3}$, and the atmospheric absorption is $\gamma=0.007~\mathrm{dB/km}$. In table.~\ref{table3}, we have evaluated the maximum detection range $R_{\mathrm{max}}$, quantum advantage $\mathcal{Q_{\mathrm{adv}}}$, and quantum range advantage $\mathcal{R_{\mathrm{adv}}}$ for these systems using equations \eqref{R_max}, \eqref{q_advantage1}, and \eqref{Range_advantage}, respectively. 
We predict that $R_{\mathrm{max}}$ for the system is proposed by Barzanjeh \textit{et. al.} \cite{barzanjehExperimentQuantumRadar2020} is greater than others, and can identify targets in the maximum range of $1.040~\mathrm{km}$, which is $1.32$ times greater than an ideal classical noise radar with the same features. 
However, the system that is implemented by Livreri \textit{et. al.} shows the highest range advantage $\mathcal{R_{\mathrm{adv}}}$ than others, which originates from the lower $N_s$ in this system. 
Note that while the QTMS radar that is realized by Luong \textit{et. al.} is faint in both maximum detection range and range advantage compared to the two other systems, but the integration time in this system is very short compared to the others, which is closest to the field applications of radar.

\subsection{Our proposed system parameters for QTMS radar for practical long-range applications}

Now, inspiring from these three systems,  we propose a feasible CW QTMS radar which is appropriate for detecting medium/small UAVs at urban distances, and shows substantial range advantage over any analogous CW classical noise radars. 
The features of our proposed system is given in the last column of table~\ref{table3}. 
In our system, we suggest to utilize the novel JTWPA implemented in \cite{liveri202210GHz} as the entangled microwave photon-pair source that generates signal and idler fields with frequencies $f_s=f_i= 9~\mathrm{GHz}$ with $N_s=0.05$ photon per mode on average. Based on \cite{zorin2019flux}, we considered the signal bandwidth as $B=9~\mathrm{GHz}$. We consider that the total gain at transmitter to be $77~\mathrm{dB}$, which is same as one has been taken in \cite{barzanjehExperimentQuantumRadar2020}. 
We also suggest to set the integration time as $\tau_{\mathrm{int}}=500~\mathrm{ms}$, which is suitable for detecting medium UAVs with maximum velocity of about $20-30~\mathrm{m/s}$. 
Such a proposed system by us enables to detect targets at ranges up to \textbf{$R_{\mathrm{max}}=1.42~\mathrm{km}$,} which is $2.14$ times greater than a classical noise radar with the same features.

For more intuition, for the case of the small UAV detection, we evaluate the maximum detection range of our proposed QTMS radar for different values of integration time $\tau_{\mathrm{int}}$ and false alarm probabilities, $P_{\mathrm{fa}}$, in Fig.~(\ref{fig8}). This figure shows that when $ P_{\mathrm{fa}}=10^{-5}$ (track mode), our system can probe targets at ranges up to \textbf{$55~\mathrm{m}$} when the integration time is set to $\tau_{\mathrm{int}}=1~\mathrm{ms}$, while the maximum detection range increases to  \textbf{$1.25~\mathrm{km}$}  when $\tau_{\mathrm{int}}=500~\mathrm{ms}$. On the other hand, when $P_{\mathrm{fa}}=0.5$ (early-alarm mode), the maximum detection range of our system is \textbf{$113~\mathrm{m}$} for $\tau_{\mathrm{int}}=1~\mathrm{ms}$, while this increases to \textbf{$2.52~\mathrm{km}$} for $\tau_{\mathrm{int}}=500~\mathrm{ms}$. 
In agreement with ours, it has been recently shown \cite{rangeEquation2023} that the quantum illumination target detection can be applied in long range, however, the authors have not considered the atmospheric loss in their approach which is very important in practical applications. Furthermore, in contrast to us, they have not obtained an explicit quantum range equation versus engineering parameters such as powers, threshold SNR, and so on. 
Note that this is crucial to link the connection between the direct detection and noise radars to better interpret what happens in the entangled quantum radars from the engineering point of view.

\subsection{Discussion and Challenges Regarding to QTMS Radars} 

To now, we have comprehensively shown the practical aspects of QTMS radars and the possibility of implementation of a km-range QTMS radar. Let us now address two major challenges regarding to the operation of QTMS radars:
	\begin{enumerate}
		\item
		Since the generated photons pair rate is low, the important question may arises can be put forwarded as: does the receiver antenna receive any photons during the integration time!?
		\item
		Does any entanglement exist between the signal-idler pair when the signal travels through the free space, interact with the target and return to the receiver antenna?
	\end{enumerate}

	\begin{figure*}
		\centering
		\subfloat[] 
		{\includegraphics[width=8cm]{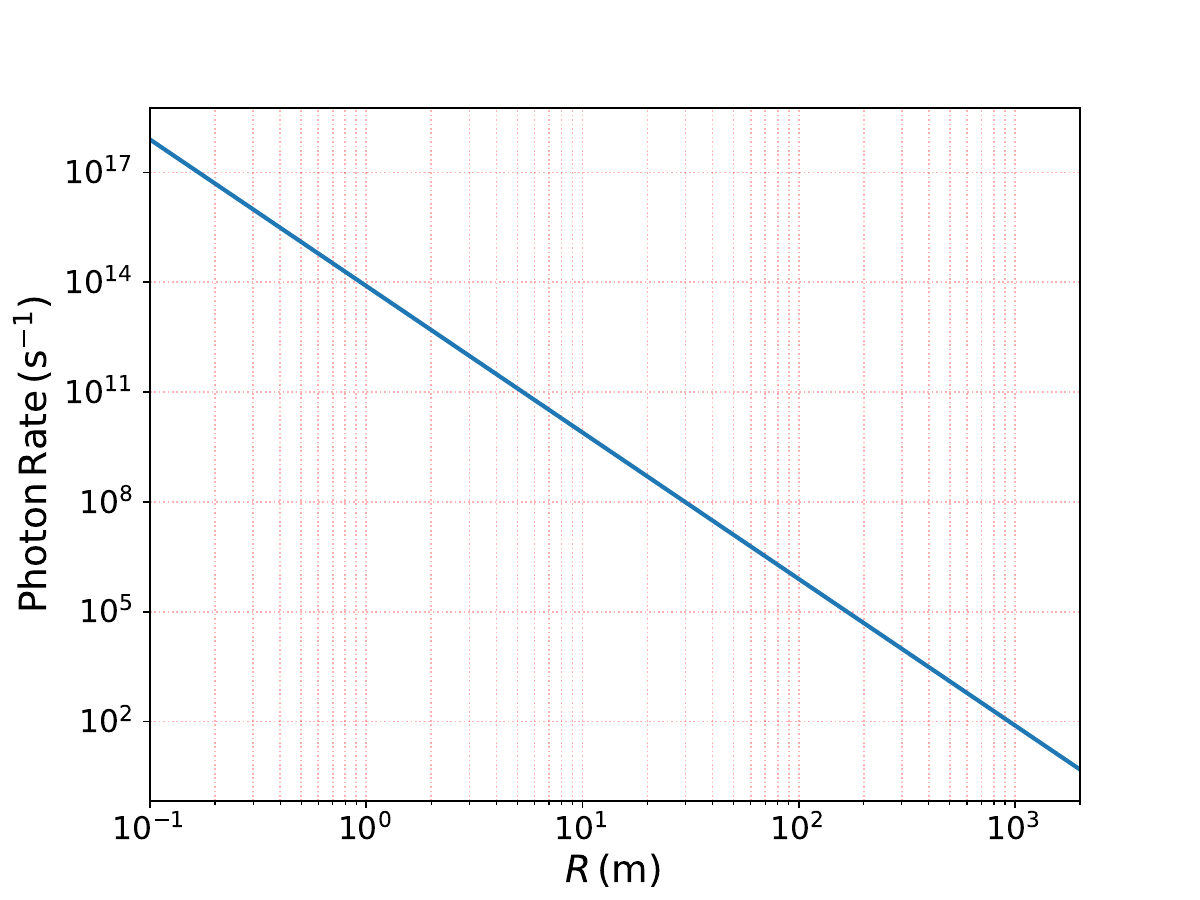}}\quad
		\subfloat[]
		{\includegraphics[width=8cm]{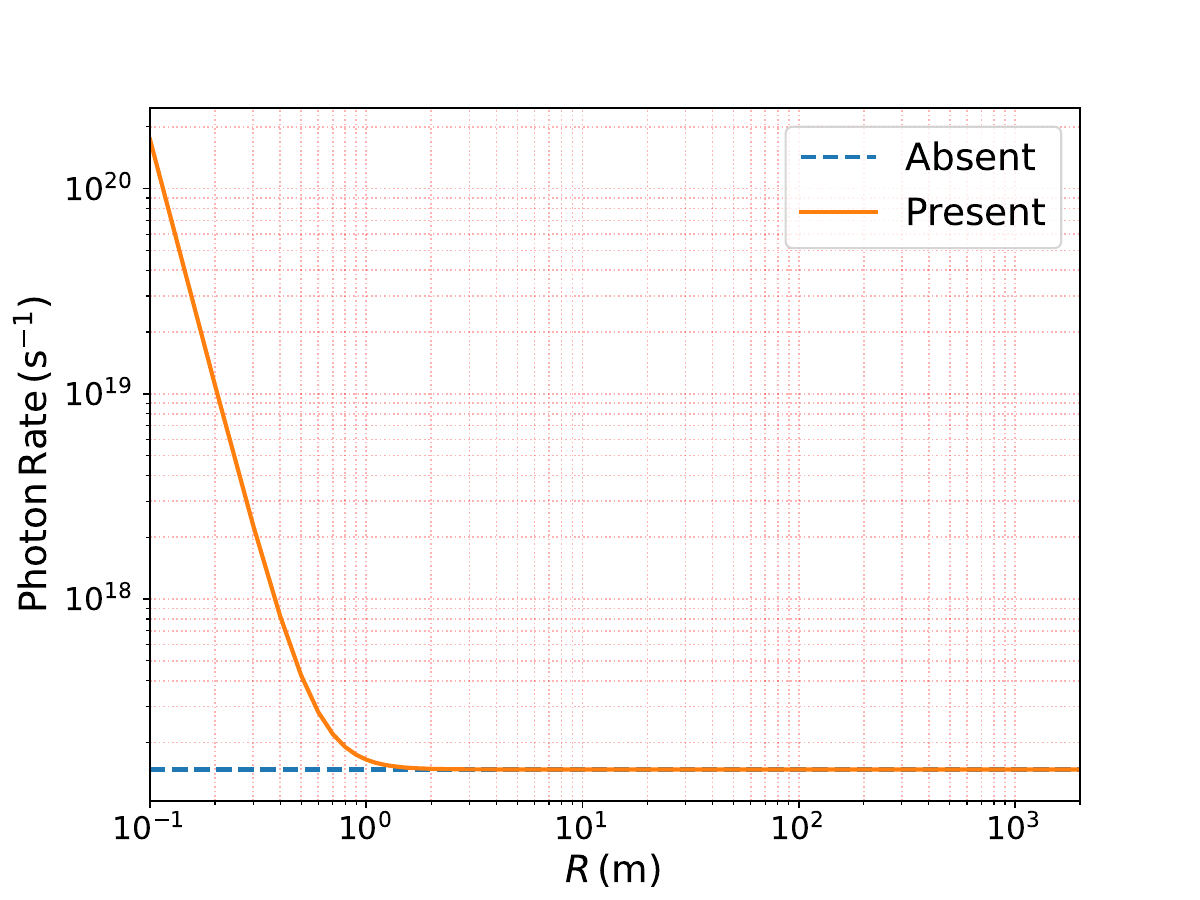}}
		\caption{(Color online) (a) The received signal-photon rate at the receiver without any noise contribution, $R_\mathrm{s}^\mathrm{corr.}$, versus the target range $R$. (b) The rate of the added noise photons due to the environmental thermal noise and added noises of the receiver and transmitter amplifiers at the receiver $R_\mathrm{n,s}^\mathrm{add}$ as a function of target range $R$ when the target is present (orange-solid) and absent (blue-dashed),, with $G_\mathrm{s}^\mathrm{det}=17 \mathrm{dB}$, $N_\mathrm{n,s}^\mathrm{tr}=9.97$, $N_\mathrm{n,s}^\mathrm{det}=3\times10^5$. Note that the total received signal photon rate is the sum of these two contributions in (a) and (b). Other parameters are same as the last column in table \eqref{table3}}.
		\label{fig10}
	\end{figure*}

	To clarify the first challenge, it is worth to mention that although the generated rates is low, but there are two stages of amplification in the system, one in transmitter and another in receiver, which substantially boosts the signal and idler power (see appendix A). According to \eqref{p_rec_signal}, if the source generates $N_s$ signal (or idler) photons, then the signal-photon number at the detector is $N_\mathrm{s}^\mathrm{corr.}=\eta(R)G_\mathrm{s}^\mathrm{amp}N_s$ where $G_\mathrm{s}^\mathrm{amp}$ is the total amplification gain on the signal mode at the transmitter as well as at the receiver (see appendix A). 
	Figure \ref{fig10}(a) demonstrates $N_\mathrm{s}^\mathrm{corr.}$ as a function of target range $R$.  The parameters considered in this figures are as the last column of table \eqref{table3}, and the amplification gain at the receiver is considered as $G_\mathrm{s}^\mathrm{det}=17 \,\mathrm{dB}$ just like Ref.~\cite{barzanjehExperimentQuantumRadar2020}. 
	As is clear in this figure even at the km-range the received signal-photon rate is substantial which implies that in the different integration times associated to defined different scenarios is enough for processing.
Note that the arrived noise photons at the receiver, $N_\mathrm{n,s}^\mathrm{det}$, is different when the target is present compare to when it is absent; when the target is present ($\eta\ne 0$), there is three contributions in the signal added noise at the receiver: environmental noise, amplification noises at the transmitter and receiver (see appendix A). 
 However, the signal added noise at the detection when the target is absent ($\eta=0$) consists of two terms, the environmental noise and the amplification noise at the receiver (see appendix A). Figure \ref{fig10}(b) demonstrates the total signal added noise at the receiver for both target presence and absence. 
 In this figure, the difference between the signal added noise is obvious at low target ranges. By comparing the received signal photon rate and signal added noise photon rate in figure \ref{fig10}(a) and (b), one observe that $R_\mathrm{s}^\mathrm{corr.}\ll R_\mathrm{n,s}^\mathrm{add}$. 
 From the viewpoint of direct detection radars, it is impossible to identify target (if present) under this condition; However, in QTMS radars, the detection take place by evaluating the correlation between the total received signal (signal+added noise) and the retained idler, which does not need the condition $R_\mathrm{s}^\mathrm{corr.} > R_\mathrm{n,s}^\mathrm{add}$ which shows the advantage of the correlated radars (both classics an quantum version) respect to the conventional radar. 
 Moreover, as is evidence in figure \ref{fig10}(b), the reduction of the signal added noise when the target is present originates to the fact that the transmitter added noise is propagated through the free space in a same manner as the radar signal, and therefore experiences a $1/R^4$ attenuation.

 Here, it is worthwhile to note that keeping the JTWPA source in cryogenic temperatures and extracting the signal-idler pairs out of the cryostat is a challenging task. In other words, the combining cryogenic operations of the microwave entangled source and of the single photon detectors, with extracting the signal out of the cryostat, launch it in free space and then retrieve it back in the cryostat are difficult process. 
 This needs to develop an efficient design of coplanar antenna for propagating entangled signals in open-air (see Ref.~\cite{gonzalez2022coplanar}).

Let us back to the second challenge regarding to the entanglement between the detected received signal and retained idler. 
To address this, one should evaluate the entanglement measures such as logarithmic negativity \cite{barzanjehQIlluminationOMS2015}, Duan-Zoller criteria \cite{Duan2000}, and Simon's separability \cite{lanzagorta2015low} before and after the detection of the signal and idler modes. 
Among them, we choose to evaluate the Simon's separability criteria. As demonstrated in appendix B, we showed that the generated signal and idler modes at the source which are represented by \eqref{SPDSstate} are inseparable, i.e., entangled, for any $N_s>0$. 
However, the entanglement between these two modes vanishes after the signal amplification, free-space propagation, and detection at the receiver (see appendix B). Therefore, the question that immediately arises is: If the quantum entanglement is so fragile so that it immediately vanishes when the signal propagates through the free space, then, where does the advantage of QTMS radars over classical noise radars come from? The answer is that although the initial entanglement completely have been destroyed when the signal propagates through free-space channel, but the absence of entanglement does not necessarily imply to classicality \cite{Weedbrook2016discord}. Indeed, the resilience of quantum entanglement in QTMS radars against environmental disturbances should explained by quantum discord or quantum correlation reconstruction as quantum resource. Quantum discord is a more resilient, and it is responsible for maintaining the benefits of entanglement in noisy channels \cite{Weedbrook2016discord}. Quantum discord is defined as the difference between the total correlations within a quantum state, as measured by the quantum mutual information, and its classical correlations, corresponding to the maximal randomness which can be shared by two parties by means of local measurements and one-way classical communication \cite{Devetak2014discord, Pirandola2014discord} (for more details see Appendix.~C). In other word, the remained quantum correlation due to the entanglement is responsible for the advantage entangled target detection over classical detection.


\section{Summary, conclusion remarks and outlooks} \label{sec6.Results and Discussion}

In this work, we addressed the performance of quantum noise radars based on TMSV states in terms of their maximum detection range by formulating the entangled quantum range equation in the microwave band. 
To this, we have started from the classical direct detection radar, in which the received signal power is the detector function, and obtained the maximum detection range for it by utilizing the detection condition in Sec.~(\ref{sec.introduction}). Then, we have shown that the maximum range of the classical direct detection radar increases by strengthening the power of the transmitted signal, and by improving the efficiency of the radar detector, which is equivalent to lowering the threshold of the signal-to-noise ratio for target detection, i.e., $\mathrm{{SNR}_{th}} $ (see Fig.~(\ref{fig2})). 
Also, we have shown that the development of microwave single-photon detectors with high efficiencies in the near future will cause a revolution and significant improvement in the $ \mathrm{{SNR}_{th}} $ which may lead to a new type of radar, i.e., single-photon radar, analogous to the single-photon LiDARs.

In Sec.~\ref{sec3.NoiseRdar}, we reviewed the idea of noise radars, in which a pair of correlated microwave signals is utilized instead of a single signal to detect the target. In this section, we studied the range equation of noise radar more complete compared to \cite{balaji2022PerformancePrediction,rangeEquation2023,bischeltsrieder2024engineering}, which utilized the classical radar range equation, as well as, they did not consider the effect of atmospheric absorption.  
In addition to considering these effects in our formulation, our novelty in this section was the definition of an effective threshold-SNR for noise radars (see \eqref{SNR_th_NR}) and derivation of an explicit expression for the maximum range of noise radars that has the same form as the one we obtained for direct detection radars (see \eqref{rangeEqDirectFinal}) and \eqref{rangeEqNR_final}). 
Based on this analogy, one can see a noise radar as a direct detection radar with reduced-threshold SNR. Finally, by obtaining the quantum entangled microwave range equation, we compared the performance of a quantum noise radar based on TMSV state (entangled states), and a classical noise radar with two IC states in Sec.~\ref{sec4.QuantumNoiseRadar}. 
We showed that the entangled signals in TMSV state has more correlations compared to those prepared in an identical coherent states, especially at the low powers (see Figures \ref{fig6} and \ref{fig7}). 
Then, we applied our formulation to the three outstanding experimental realizations of QTMS radars in Sec.~\ref{sec5.experiment} and estimated their maximum detection range $R_{\mathrm{max}}$, quantum advantage $\mathcal{Q_{\mathrm{adv}}}$, and quantum range advantage $\mathcal{R}_{\mathrm{adv}}$. 
Moreover, we proposed a feasible parameters for QTMS radar system in which a JTWPA is utilized as an entangled microwave photon-pair source that generates photon pair with degenerate frequencies $9~\mathrm{GHz}$ (which demonstrated experimentally in \cite{liveri202210GHz}), and bandwidth of $\mathrm{9~GHz}$ \cite{zorin2019flux}. 
We showed that this system can recognize targets with $\mathrm{RCS}=0.1~\mathrm{m}^2$ such as small UAVs in maximum range in order of \textbf{$1.25~\mathrm{km}$} when working in the track mode, and in range up to $2.5~\mathrm{km}$ when working in the early alarm mode which is suitable for detecting UAVs in urban distances.

As outlooks for improving the QTMS radars in order to be applied from the engineering and practical point of view, one can address to
\begin{itemize}
	\item
	Engineering of the microwave entangled sources (based on the electro-optomechanics (EOM) [for review on optomechnics see Ref.~\cite{aspelmeyer2014OMS}], superconducting or other platforms such as NV or SPDC) with `increased bandwidth' as much as possible, because this plays a crucial role in the enhancement of the maximum detection range as well as the quantum range advantage;
	\item
	Developing the photonics radar platform \cite{ghelfiPhotonicRadar1,PhotonicRadar2WiePan,photonicRadar3}. Since it seems that if the optomechanical-based microwave quantum entangled radars will be developed, its platform is compatible to the photonics radar;	
	\item 
	`Entanglement enhancement' and `room-temperature' operation of microwave photon-pair sources based on EOMs or superconducting systems through quantum control (see Refs.~\cite{painter2023,vitaliCoverter2023,marinCooing2023Room,microwave_Photon_Emission_Superconducting_Circuits,kippenbergRoomTemperaure2024}). This enhances the SWaP (size, wight, and power consumption) and cost of the system substantially;
	\item 
	Developing single-photon microwave detectors for replacing with the current receiver antennas in order to substantially improve the `threshold-SNR' of entangled noise radar;
	\item 
	Investigating the `other platforms' such as nonlinear bulk/integrated optics crystals or NV-centers which works at room-temperature for generation of entanglement in GHz-band in order to enhance the SWaP and cost of the system by omitting the cryostat.
\end{itemize}

\section*{Acknowledgment}
AMF would thank ICQT. Both authors also thank reviewers for their constructive comments which substantially enhance the manuscript.  
\textit{Contributions}. AMF as the corresponding author and director of the quantum sensing $ \& $ metrology Lab/group at ICQT has defined and led this project.  
All calculations have been done by HA and rechecked and interpreted by AMF. All numerical calculations and graphs have been plotted by HA. The subjects of the quantum antenna or single-photon microwave detectors for improving the threshold-SNR have been introduced and developed by AMF. The effective-SNR has been defined and developed by HA and has been interpreted by AMF. Experimental parameters have been determined and extracted by both authors. Both authors have contributed to the discussion and interpretation of the results as well as writing, editing and revising the manuscript. The idea of investigating the received photon number and Simon criterion are put forwarded by AMF, calculated by HA and then re-checked by AMF. Both authors contributed to answer to the reviewers.

\appendix

\section{A: Squeezing origination of microwave QTMS state in JPA}
It is well-known that in the optical domain, the entangled Gaussian states can be generated by SPDC, a 3-wave mixing process in NLCs in which a pump photon is converted through second-order nonlinear interaction with quantum vacuum fluctuations into a pair of signal-idler photons respecting energy-momentum conservation law. Since signal-idler biphoton is created from the same vacuum, they both share entanglement between each other stronger than any classical correlation.
On the other hand, in the microwave band, phase-preserving JPAs, as a quantum-limited amplifiers that operate in 3-10GHz band with signal-idler frequency which can be adjusted in GHz-band, are responsible for the microwave entanglement generation via the well-known 4-wave mixing process \cite{wilsonProgress2020}. 
In a JPA, two input vacuum signal ports ($ \hat \nu, \hat \mu $) are mixed together and two amplified sideband signal and idler with power gain $ G_{\mathrm{JPA}} $ are generated. The generation of entanglement from uncorrelated vacuum noise inputs in a JPA can be modeled as \cite{wilsonProgress2020} 
\begin{eqnarray}
&& \hat a_{\mathrm{s}}=\cosh (r) \hat \nu + \sinh (r) \hat \mu^\dagger , \\
&&  \hat a_{\mathrm{i}}=\sinh (r) \hat \nu^\dagger + \cosh (r) \hat \mu,
\end{eqnarray}
where $ r $ is referred to the squeezing parameter and is related to the amplification gain power as $ G_{\mathrm{JPA}}=\cosh^2(r) $. 
It is simple to show that the signal-idler quadratures of a QTMS state is a zero-mean Gaussian thermal state with average photon number as the so-called photon mer mode $ N_s=[\cosh(2r)-1]/2 $ which immediately implies that signal-idler correlation in the Covariance matrix becomes $ \mathcal{C}_Q=\sinh(2r)/2 $ and can be greater than its classical counterpart.
Surprisingly, this quantum advantage is responsible for the noise reduction below quantum vacuum ($ 1/2 $) in generalized quadratures $ \hat I_-=(\hat I_{\mathrm{s}}-\hat I_{\mathrm{i}})/\sqrt{2} $ and $ \hat Q_+=(\hat Q_{\mathrm{s}}+\hat Q_{\mathrm{i}})/\sqrt{2} $. This leads to quadrature squeezing as
\begin{eqnarray}
&& Var[\hat I_-]=e^{-2r}/2, 
\end{eqnarray}
which has been experimentally realized up to around -12dB \cite{microwaveSquuezing2014}.

\section{B: Evaluation of Signal and Idler power in noise radars} \label{app_B}

In this appendix, we follow the approach in Ref.~\cite{barzanjehExperimentQuantumRadar2020} to evaluate the power of signal and idler at the transmitter and receiver. If the annihilation operator of the generated signal and idler at the source is $\hat{a}_\mathrm{s}$ and $\hat{a}_\mathrm{s}$, respectively, the annihilation operator of these fields after the amplification at the transmitter is then given by
	\begin{align}
		\hat{a}_\mathrm{s}^\mathrm{tr}&=\sqrt{G_\mathrm{s}^\mathrm{tr}}\hat{a}_\mathrm{s}+\sqrt{G_\mathrm{s}^\mathrm{tr}-1}(\hat{a}_\mathrm{n,s}^\mathrm{tr})^\dagger,\label{eq. A1-1-1}\\
		\hat{a}_\mathrm{i}^\mathrm{tr}&=\sqrt{G_\mathrm{i}^\mathrm{tr}}\hat{a}_\mathrm{i}+\sqrt{G_\mathrm{i}^\mathrm{tr}-1}(\hat{a}_\mathrm{n,i}^\mathrm{tr})^\dagger,\label{eq. A1-1-2}
	\end{align}
	in which $G_\mathrm{k}^\mathrm{tr}$ and $\hat{a}_\mathrm{n,k}^\mathrm{tr}$ are the amplification gain and the added noise annihilation operator at the radar transmitter, respectively, with $\mathrm{k=s}$ for the signal and $\mathrm{k=i}$ for the idler modes. The idler mode, after generation and amplification, is down-converted to a base-band and after another amplification stage, it directly detected, digitized, and maintained at the transmitter as a reference for target detection. The annihilation operator of the detected idler field is given by
	\begin{align}
		\hat{a}_\mathrm{i}^\mathrm{det}=\sqrt{G_\mathrm{i}^\mathrm{det}}\hat{a}_\mathrm{i}^\mathrm{tr} + \sqrt{G_\mathrm{i}^\mathrm{det}-1}(\hat{a}_\mathrm{n,i}^\mathrm{det})^\dagger, \label{eq. A1-2}
	\end{align}
	in which $G_\mathrm{i}^\mathrm{det}$ and $\hat{a}_\mathrm{n,i}^\mathrm{det}$ are the detection amplification gain and detection added noise annihilation operator of the idler mode. Using equations \eqref{eq. A1-1-2} and \eqref{eq. A1-2}, the mean photon number of the idler mode at the detector is obtain as
	\begin{align}
		N_\mathrm{i}^\mathrm{det}&=\braket{(\hat{a}_\mathrm{i}^\mathrm{det})^\dagger\hat{a}_\mathrm{i}^\mathrm{det}} \label{eq. A1-3-1}\\
		&=G_\mathrm{i}^\mathrm{amp}N_0+N_\mathrm{n,i}^\mathrm{add},\label{eq. A1-3-2}
	\end{align}
	in which $G_\mathrm{i}^\mathrm{amp}\equiv G_\mathrm{i}^\mathrm{det}G_\mathrm{i}^\mathrm{tr}$ is the total amplification on the idler mode, $N_0\equiv\braket{\hat{a}_\mathrm{i}^\dagger\hat{a}_\mathrm{i}}$ is the number of idler photons generated at the source, and $N_\mathrm{n,i}^\mathrm{add}$ is the number of noise photons added to the idler mode during two stages of amplification which is given by
	\begin{align}
		N_\mathrm{n,i}^\mathrm{add}=&G_\mathrm{i}^\mathrm{det}(G_\mathrm{i}^\mathrm{amp}-1)N_\mathrm{n,i}^\mathrm{tr}+(G_\mathrm{i}^\mathrm{det}-1)N_\mathrm{n,i}^\mathrm{det}\notag\\
		&+(G_\mathrm{i}-1)  \label{eq. A1-4}
	\end{align}
	where $N_\mathrm{n,i}^\mathrm{tr}\equiv\braket{(\hat{a}_\mathrm{n,i}^\mathrm{tr})^\dagger\hat{a}_\mathrm{n,i}^\mathrm{tr}}$ and $N_\mathrm{n,i}^\mathrm{det}\equiv\braket{(\hat{a}_\mathrm{n,i}^\mathrm{det})^\dagger\hat{a}_\mathrm{n,i}^\mathrm{det}}$ are the number of added noise photons through the first and second amplification stages, respectively. Therefore, it is straightforward to show that the idler power is
	\begin{align}
		P_\mathrm{i}= P_\mathrm{i}^\mathrm{corr} + P_\mathrm{n,i}, \label{eq. A1-5}
	\end{align}
	in which $P_\mathrm{i}^\mathrm{corr}=G_\mathrm{i}^\mathrm{amp}N_0hf_iB$ and $P_\mathrm{n,i}=N_\mathrm{n,i}^\mathrm{add}hf_iB$ are the power of the correlated part of the idler and the idler added noise power, respectively. This equation is equivalent to \eqref{signal_idler_power}.

	The situation for the signal mode is slightly different; it propagates through the free space and interacts with the target (if present) after the first amplification stage (which is represented by \eqref{eq. A1-1-1}). It is then received by the receiver antenna, down-converts to the base-band, amplified with another amplification stage, and finally digitized and processed to extract the target information.  If we consider the transmitter-target-receiver channel transmission as $\eta$, then the annihilation operator for the signal mode at the receiver is given by
	\begin{align}
		\hat{a}_\mathrm{s}^\mathrm{rec}=\sqrt{\eta}\hat{a}_\mathrm{s}^\mathrm{tr} + \sqrt {1-\eta}\hat{a}_\mathrm{n}^\mathrm{env}, \label{eq. A1-6}
	\end{align}
	in which $\hat{a}_\mathrm{n}^\mathrm{env}$ is the annihilation operator of the environmental noise added to the signal through the free space transmission. Note that we can model the absence of the target by setting $\eta=0$, and the presence of the target by $\eta\neq0$. After receiving the signal, down-convert to the base-band, and amplification, the signal mode can be represented by
	\begin{align}
		\hat{a}_\mathrm{s}^\mathrm{det}=\sqrt{G_\mathrm{s}^\mathrm{det}}\hat{a}_\mathrm{s}^\mathrm{rec}+\sqrt{G_\mathrm{s}^\mathrm{det}-1}(\hat{a}_\mathrm{n,s}^\mathrm{det})^\dagger, \label{eq. A1-7}
	\end{align}
	in which $G_\mathrm{s}^\mathrm{amp}$ is the amplification gain of the signal in the receiver and $\hat{a}_\mathrm{n,s}^\mathrm{det}$ is the annihilation operator of the added noise through this amplifier. Then, one can show that the mean photon number of the detected signal is
	\begin{align}
		N_\mathrm{s}^\mathrm{det}&=\braket{(\hat{a}_\mathrm{s}^\mathrm{det})^\dagger \hat{a}_\mathrm{s}^\mathrm{det}} \label{eq. A1-8-1}\\
		&=\eta G_\mathrm{s}^\mathrm{amp}N_0 + N_\mathrm{n,s}^\mathrm{add},\label{eq. A1-8-2}
	\end{align}
	in which $G_\mathrm{s}^\mathrm{amp}\equiv G_\mathrm{s}^\mathrm{det}G_\mathrm{s}^\mathrm{tr}$ is the total amplification on the signal mode, $N_0\equiv\braket{\hat{a}_\mathrm{s}^\dagger\hat{a}_\mathrm{s}}$ is the number of signal photons at the source, and $N_\mathrm{n,s}^\mathrm{add}$ is the number of noise photons added to the signal through the free space propagation and two stages of amplification, so that
	\begin{align}
		N_\mathrm{n,s}^\mathrm{add}=&\eta G_\mathrm{s}^\mathrm{det}(G_\mathrm{s}^\mathrm{tr}-1)N_\mathrm{n,s}^\mathrm{tr}+ (G_\mathrm{s}^\mathrm{det}-1)N_\mathrm{n,s}^\mathrm{det}\notag\\
		&+(1-\eta)G_\mathrm{s}^\mathrm{det}(N_\mathrm{n}^\mathrm{env}+1) + (\eta G_\mathrm{s}^\mathrm{amp}-1),\label{eq. A1-9}
	\end{align}
	where
	$N_\mathrm{n,s}^\mathrm{tr}\equiv\braket{(\hat{a}_\mathrm{n,s}^\mathrm{tr})^\dagger\hat{a}_\mathrm{n,s}^\mathrm{tr}}$ and
	$N_\mathrm{n,s}^\mathrm{rec}\equiv\braket{(\hat{a}_\mathrm{n,s}^\mathrm{rec})^\dagger\hat{a}_\mathrm{n,s}^\mathrm{rec}}$
	are the number of noise photons added at the first and second amplification stages, respectively, and
	$N_\mathrm{n}^\mathrm{env}\equiv\braket{(\hat{a}_\mathrm{n}^\mathrm{env})^\dagger\hat{a}_\mathrm{n}^\mathrm{env}}$ is the number of noise photons added to the signal mode through the free space propagation. By utilizing \eqref{eq. A1-8-2} it is easy to show that the detected signal power is given by
	\begin{align} 
		P_\mathrm{s}=\eta P_\mathrm{s}^\mathrm{corr} + P_\mathrm{n,s} \label{eq. A1.10}
	\end{align}
	in which $P_\mathrm{s}^\mathrm{corr}=G_\mathrm{s}^\mathrm{amp}N_0 hf_sB$ is the power of the correlated part of the signal, and $P_\mathrm{n,s}=N_\mathrm{n,s}^\mathrm{add}hf_sB$ is the power of the noise added to the signal mode. This equation is equivalent to \eqref{p_rec_signal}.

\section{C: Simon's Separability Criteria Evaluation}
Here, in this appendix we are going to calculate the Simon's separability criteria for the signal and idler modes before the transmission of the signal and also after returning to the receiver in  general situation by considering attention, amplification and noise. To this, we starts with the covariance matrix for signal and idler modes at the transmitter which is given by \cite{lanzagorta2015low}
	\begin{align}
		\mathcal{C}_\mathrm{s,i}^\mathrm{tr}=\frac{1}{4}\left( {\begin{array}{*{20}{c}}
				S_1 & 0 & C_q & 0 \\
				0 & S_1 & 0  & -C_q\\
				C_q & 0 & S_2 & 0 \\
				0 & -C_q & 0 & S_2
		\end{array}} \right),\label{eq. A2-1}
	\end{align}
	in which $S_1=S_2= 2N_s + 1$, and $C_q = \sqrt{N_s(N_s + 1)}$.
	For evaluating the correlation between the detected signal and the retained idler, we begin by \eqref{eq. A1-2} and \eqref{eq. A1-7}:
	\begin{align}
		\hat{a}_\mathrm{i}^\mathrm{det}=&\sqrt{G_\mathrm{i}^\mathrm{amp}}\hat{a}_i + \sqrt{G_\mathrm{i}^\mathrm{det}(G_\mathrm{i}^\mathrm{tr}-1)}(\hat{a}_\mathrm{n,i}^\mathrm{tr})^\dagger \notag \\
		&+\sqrt{G_\mathrm{i}^\mathrm{det}-1}(\hat{a}_\mathrm{n,i}^\mathrm{det})^\dagger \label{eq. A2. 2}\\
		\hat{a}_\mathrm{s}^\mathrm{det}=&\sqrt{\eta G_\mathrm{s}^\mathrm{amp}}\hat{a}_\mathrm{s} + \sqrt{G_\mathrm{s}^\mathrm{det}(1-\eta)}\hat{a}_\mathrm{n}^\mathrm{env} \notag\\
		&+\sqrt{G_\mathrm{s}^\mathrm{det}(G_\mathrm{s}^\mathrm{tr}-1)}(\hat{a}_\mathrm{n,s}^\mathrm{tr})^\dagger + \sqrt{G_\mathrm{s}^\mathrm{det}-1}(\hat{a}_\mathrm{n,s}^\mathrm{det})^\dagger. \label{eq. A2. 3}
	\end{align}
	Using the definition of the in-phase and quadrature voltages of the field as $\hat{I}_\mathrm{k}=(\hat{a}_\mathrm{k}+\hat{a}_\mathrm{k}^\dagger)/\sqrt{2}$ and $\hat{Q}_\mathrm{k}=(\hat{a}_\mathrm{k}-\hat{a}_\mathrm{k}^\dagger)/i\sqrt{2}$ with $\mathrm{k=s, \, i}$ for signal and idler modes, respectively, then the covariance matrix of the detected signal and idler modes read
	\begin{align}
		\mathcal{C}_\mathrm{s,i}^\mathrm{det}=\frac{1}{4}\left( {\begin{array}{*{20}{c}}
				S_1' & 0 & C_q' & 0 \\
				0 & S_1' & 0  & -C_q'\\
				C_q' & 0 & S_2' & 0 \\
				0 & -C_q' & 0 & S_2'
		\end{array}} \right), \label{eq. A2. 4}
	\end{align}
	in which
	\begin{align}
		S_1'=&2\big[\eta G_\mathrm{s}^\mathrm{amp} N_\mathrm{s} + G_\mathrm{s}^\mathrm{det}N_\mathrm{n}^\mathrm{env} + G_\mathrm{s}^\mathrm{det}(G_\mathrm{s}^\mathrm{tr}-1)(N_\mathrm{n,s}^\mathrm{tr}+1) \notag\\
		&+(G_\mathrm{s}^\mathrm{det}-1)(N_\mathrm{n,s}^\mathrm{det}+1)\big]+1, \label{eq. A2. 5}\\
		S_2'=&2\big[G_\mathrm{i}^\mathrm{amp}N_\mathrm{s} + G_\mathrm{i}^\mathrm{det}(G_\mathrm{i}^\mathrm{tr}-1)(N_\mathrm{n,i}^\mathrm{tr}+1)  \notag\\
		&+(G_\mathrm{i}^\mathrm{det}-1)(N_\mathrm{n,i}^\mathrm{det}+1)\big]+1 \label{eq. A2. 6} \\
		C_q'=& \sqrt{\eta G_\mathrm{s}^\mathrm{amp}G_\mathrm{i}^\mathrm{amp}}C_q, \label{eq. A2. 7}
	\end{align}
	in which $N_\mathrm{s}$, $N_\mathrm{n,s}^\mathrm{tr}$, $N_\mathrm{n,s}^\mathrm{det}$, $N_\mathrm{n,i}^\mathrm{tr}$, and $N_\mathrm{n,i}^\mathrm{det}$ are defined in appendix A. It is worth noting that when there is no amplification, i.e., $G_\mathrm{k}^\mathrm{det}=G_\mathrm{k}^\mathrm{tr}=1$, the covariance matrix given in \eqref{eq. A2. 4} reduces to that reported in references \cite{tan2008quantum, lanzagorta2015low} for QTMS radar with no amplification. 
	
	As mentioned in \cite{lanzagorta2015low}, the entanglement between the signal and idler mode can be evaluated through the so-called Simon's separability criteria as
	\begin{align}
		f \ge 0,
	\end{align}
	in which $f$ is defined as Simon parameter, and for the signal and idler modes represented by the covariance matrix of the form \eqref{eq. A2-1} or \eqref{eq. A2. 4} is defined as
	\begin{align}
		f \equiv (S_1 S_2-C_q^2)^2 - (S_1^2 + S_2^2 + 2C_q^2) + 1. \label{eq. A2. 8}
	\end{align}
	Using this equation, it is straightforward to show that $f$ at the source is given by
	\begin{align}
		f_\mathrm{source}=-16N_s(N_s+1), \label{eq. A2. 9}
	\end{align}
	which is negative for any $N_\mathrm{s}>0$. Therefore, it is concluded that the generated signal and idler modes are inseparable or entangled. However, using \eqref{eq. A2. 5}-\eqref{eq. A2. 7} one can obtain $f$ for the detected signal and idler modes as
	\begin{align}
		f_\mathrm{det} = (S_1'S_2' - C_q'^2)^2 - (S_1'^2 + S_2' + 2C_q') + 1. \label{eq. A2. 10}
	\end{align}
	by considering $G_\mathrm{s}^\mathrm{tr}=G_\mathrm{i}^\mathrm{tr}=77~\mathrm{dB}$, $G_\mathrm{s}^\mathrm{det}=G_\mathrm{i}^\mathrm{det}=17~\mathrm{dB}$, and the parameters as in the last column of table \eqref{table3}, it is concluded that $f_\mathrm{det}\gg 1$ for any $R>0$. Therefore, the detected signal and idler modes are not entangled.

\begin{IEEEbiography}[{\includegraphics[width=1in,height=1.25in,clip,keepaspectratio]{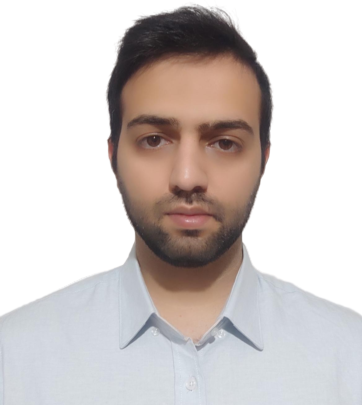}}]{H. Allahverdi} received his M.Sc. in photonics from Shahid Beheshti University (SBU) in 2020. He is currently pursuing his Ph.D. in photonics at the Laser and Plasma Research Institute (LAPRI) at SBU. 
His research includes quantum sensing and metrology, quantum optomechanics, and atom interferometry. In addition to theoretical studies, he is skilled in the design and implementation of fiber optical setups, ultra-high vacuum systems, and electro-optic light modulators. He has joined to MotazediFard group from 3-years ago for part-time research projects on quantum remote sensing.
\end{IEEEbiography}

\begin{IEEEbiography}[{\includegraphics[width=1.1in,height=2in,clip,keepaspectratio]{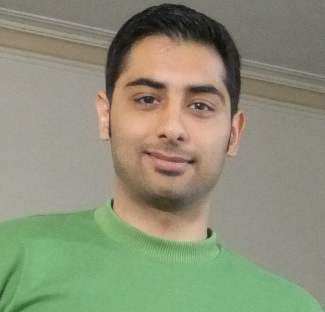}}]{Ali Motazedifard} born in 1987 received his Master's degree of physics in experimental optical metrology in 2012 at the University of Tehran and his PhD in Quantum Optics in the field of Quantum Control of linear or nonlinear electro-opto-mechanical bare- and hybrid-systems assisted by the cold atoms and BEC for sensing applications, generating non-classical features for mechanics and light, or communications at the University of Isfahan in 2018. He ranked 3rd/5th of the physics Olympiad in the Tehran/IRAN in 2008. He was honored to award National Alimohammadi Prize in 2020 for the best PhD thesis in physics in IRAN. 
In 2014-2015, he was a visiting researcher in the Quantum Optics group at the University of Camerino in Italy. 
During 2009-2017, He worked professionally as a part-time experimental researcher, respectively, at Kahroba Company and R$\&$D optics- 
and microelectronics-center of ACECR-Sharif University of Technology on the micro-lithography, sputtering, and manufacturing optical measurement devices using interferometry, Fresnel diffractometry/scattering/refractometry 
and Moire deflectormetry. 
He was awarded a Fellowship on behalf of Iran National Elites Foundation (INEF) to be a visiting researcher scientist for 2-years at the Institute for Research in Fundamental Sciences (IPM). Since 2018, he has been working as a senior experimental researcher at the Iranian Center for Quantum Technologies (ICQT) and contributing extensively to projects supported by it. From 2020, He has been director and leader of the Quantum Sensing and Metrology Group and Lab at ICQT. He was the leader, manager and also the principal experimental researcher of the most important projects that have been done for the first time in IRAN such as SPDC-based optical entanglement generation, attenuated single-photon generation, quantum diagnosis of DNA, secure free-space polarization-entanglement-based quantum cryptography and communication links up to 6km, construction of the optocoupler-based quantum random number generator and FPGA-based sub-nanosecond time-tagger for LiDAR applications, and AI-improved quantum-structured-noise computational Ghost imaging. In addition, recently, his research group activities have been focused on some long-term experimental projects: quantum remote sensing including quantum LiDAR and Radar, Cold-atom-based sensors, and theoretical development of the quantum sensors and radiation sources based on the cavity optomechanical systems.
\end{IEEEbiography}

\EOD

\end{document}